\documentclass[aps,prc,twocolumn,showpacs,floatfix,amsmath,amssymb,preprintnumbers]{revtex4}
\usepackage{graphicx}
\usepackage{amsmath}
\usepackage{color}
\graphicspath{{figs/}}

\newcommand{\rs}[1]{_{\mbox{\tiny #1}}}
\newcommand{\bm}[1]{\mbox{\boldmath $#1$}} 

\newcommand{\fctd}{f_{\Delta}} 
\newcommand{\facomega}{f_{\omega}} 
\newcommand{\numnuclp}{N_{\rm p}} 
\newcommand{\numnuclo}{N_{\rm o}} 
\newcommand{\numnucls}{N_{\rm s}} 
\newcommand{\numnuclwdp}{N_{\rm p}^{\mbox{\tiny DM}}} 
\newcommand{\numnuclwdo}{N_{\rm o}^{\mbox{\tiny DM}}} 
\newcommand{\nosc}{N_{\rm osc}} 
\newcommand{\rp}{R_{\mathrm{p}}}
\newcommand{\rpp}{R'_{\mathrm{p}}}
\newcommand{\rppp}{R''_{\mathrm{p}}}
\newcommand{\ropp}{R''_{\mathrm{o}}}
\newcommand{\rspp}{R''_{\mathrm{s}}}

\begin{document}

\title{
Nuclear prolate-shape dominance with the Woods-Saxon potential
}

\author{Satoshi Takahara}
\affiliation{Kyorin University, School of Medicine, Mitaka, Tokyo 181-8611, Japan}
\author{Naoki Tajima}
\affiliation{Department of Applied Physics, University of Fukui, 3-9-1 Bunkyo, Fukui 910-8507, Japan}
\author{Yoshifumi R. Shimizu}
\affiliation{Department of Physics, Graduate School of Science, Kyushu University, Fukuoka 812-8581, Japan}

\date{\today}

\begin{abstract}
We study the prolate-shape predominance of the nuclear ground-state deformation
by calculating the masses of more than two thousand even-even nuclei
using the Strutinsky method, modified by Kruppa, and improved by us.
The influences of the surface thickness of the single-particle potentials,
the strength of the spin-orbit potential,
and the pairing correlations are investigated by varying the parameters
of the Woods-Saxon potential and the pairing interaction.
The strong interference between the
effects of the surface thickness and the spin-orbit potential is
confirmed to persist for six sets of the Woods-Saxon potential parameters.
The observed behavior of the ratios of prolate, oblate, and spherical nuclei
versus potential parameters are rather different in different mass regions.
It is also found that the ratio of spherical nuclei increases
for weakly bound unstable nuclei.
Differences of the results from the calculations with the
Nilsson potential are described in detail.
\end{abstract}

\pacs{
 21.10.Ft, 
 21.10.Gv, 
 21.10.Ky, 
 21.10.Pc  
}

\maketitle

\section{INTRODUCTION}
It has long been known that most of the deformed nuclei have prolate rather than
oblate shapes~\cite{BM75}.
This {\it prolate-shape dominance} is the subject of this paper,
in which we examine the effects of several important factors
to increase or decrease such preference for prolate shapes.

The properties of a nucleus are completely determined by the
Hamiltonian for quantum mechanical many-body systems consisting of
nucleons.  From this point of view, a possible direction of research
for the origin of the prolate-shape dominance is to change
artificially each term of the Hamiltonian and examine its effect on
the prolate-shape dominance.  The most successful outcome of such a study
would be a discovery of a direct correspondence between the
prolate-shape dominance and a specific term of the elementary
nucleon-nucleon interaction, in an analogous fashion as the tensor
force causes a mixture of the $d$ wave in the wavefunction of a deuteron.
However, nobody knows whether there exists such a simple relation.

In mean-field approximations all the influences of the many-body
Hamiltonian on the nuclear shape are conveyed through the mean-field (one-body)
Hamiltonian.
Hence, one can seek the origin of the prolate-shape dominance in the properties
of the mean field, instead.
If one also adopts the macroscopic-microscopic method~\cite{Strt},
which uses only the single-particle level density,
one can determine the nuclear shape starting from the mean-field Hamiltonian
without referring to the underlying many-body Hamiltonian.
This is the choice of this paper.
One can reasonably expect that the linkage from the mean-field
Hamiltonian to the nuclear shape may be simpler than those from the
many-body Hamiltonian.
Moreover, the former linkage has an established route to understand in terms of
classical periodic orbits~\cite{Frisk90,Ari12}, while the latter has none presently.

We have been seeking the origin of the prolate-shape dominance in the
one-body potentials in the macroscopic-microscopic framework.
In Ref.~\cite{TS01}, we employed the Nilsson potential~\cite{NTS69} and
calculated the ratio of prolate-shape nuclei among $\sim$1800 even-even nuclei
as a function of the strengths of the $l^2$ term and the spin-orbit term.
We found that
the ratio of prolate nuclei oscillates versus the strength of the
spin-orbit term when the $l^2$ term has the standard strength.
This oscillation disappears when the $l^2$ potential is weaker.
The amplitude of the oscillation is quite large.
The ratio of prolate-shape nuclei oscillates
between $\sim$40\% (at $\pm$40\% strengths of the standard)
and $\ge$80\% (at $\pm$100\% and 0\% strengths).
This result indicates a strong interference between the effects of
the two terms of the potential.

However, this interference might be exaggerated because of the affinity
of the two terms in the sense that both of them include the orbital
angular momentum operator $l$. The $l^2$ term is an approximation to the
radial profile of the central potential like a square well.
In a letter~\cite{TOST11}, we replaced the Nilsson potential with
the Woods-Saxon (WS) potential to eliminate the
$l^2$ potential. As a result, we still find the oscillation,
although its amplitude was decreased by a factor $\sim \frac{1}{3}$.

Employment of the WS potential
required various modifications of the method~\cite{TST10}.
First, the WS potential has a continuum
spectrum, which cannot be treated by the standard Strutinsky
method. This problem was solved by the Kruppa's prescription~\cite{Kruppa98}
to subtract the level density of the free-particle spectrum.  
Concerning the remaining problem of the plateau condition,
we invented the reference density method~\cite{TST10} 
to make the shell energy insensitive to the smearing width.
Second, the BCS equation with a constant pair-scattering
matrix element $G$ also suffers from the continuum spectrum. We
extended the Kruppa's prescription to the paring to subtract the
contributions from free-particle spectrum in the BCS equation.
This treatment also improves the calculation of the nuclear radius~\cite{OSTT10}.
Third, available parameter sets of the
WS potentials are not consistent with the macroscopic part
especially for nuclei far from the $\beta$-stability line.
We presented a method based on the Thomas-Fermi
approximation to adjust the depth of the central potential to reproduce
the drip lines predicted by the macroscopic energy.

In this paper, we show and discuss the results of the calculations
with the WS potential in much more detail than in the
letter~\cite{TOST11}.
We try not a single but six parameter sets of the WS potential.
We confirm the convergence of the results versus the size of the
truncated oscillator basis for the diagonalization of the
single-particle Hamiltonian.
We show that different definitions of the ratio of prolate-shape
nuclei do not change the conclusions.
We also consider the effects of changing the strength of the pairing
interaction, which we studied so far only for the Nilsson potential in
Ref.~\cite{TSS02}.
We show the distribution of prolate and oblate nuclei in the nuclear
chart for various modified potentials and confirm the independence of
the main conclusions from the choice of potentials.

In the rest of the introduction, we survey the earlier works.

For the anisotropic harmonic oscillator potential, which is
the most basic approximation to the one-nucleon potential,
prolate (oblate) shapes tend to
have lower energies than oblate (prolate) ones
when the major shell is less (more) than half filled
because the particles (holes) in the low $\Omega$ orbitals
drives strongly the nuclear shape toward prolate (oblate) direction.
Here, $\Omega$ is the projection of the single-particle angular momentum 
onto the symmetry axis of the mean-field potential. 
This suggests an approximately equal number of prolate and oblate nuclei~\cite{BM75}.
Indeed, our quantitative estimation~\cite{TS01} showed that
the ratio of prolate nuclei among deformed nuclei is 55\% for a harmonic oscillator
potential, while the experimental ratio of prolate nuclei seems much larger.

The earliest attempt to explain the prolate-shape preference is a work
by Lemmer and Weisskopf~\cite{LW61}.
A term $\propto r^4$ was added to the harmonic oscillator potential
to steepen the wall. Then it overrode the normal tendency of the oscillator
potential to deform into an oblate shape for a more-than-half-filled major shell.
It suggests that the origin of the prolate-shape dominance
is the radial profile of the potential.

However, there was also an opinion that the origin is the spin-orbit
potential.  For example, Ref.~\cite{BM75} states that it is the spin-orbit
potential which breaks the even situation by weakening the
oblate-shape shell effect in $sd$ shell nuclei.

The truth has turned out~\cite{TS01} that both of the two salient
features of the nuclear single-particle potential,
the square-well like radial profile and the spin-orbit potential, play essential roles to
give rise to the prolate-shape dominance. The dominance can be equally
reproduced with and without the spin-orbit potential but it does
not deny the participation of the spin-orbit potential,
because half-strength spin-orbit potential destroys the dominance.

The relation between potential and spectrum can be understood
in terms of classical periodic orbits.
Putting aside the spin-orbit potential,
Frisk analyzed classical periodic orbits in an ellipsoidal cavity and
found that the radial dependence is an origin of the prolate-shape
dominance~\cite{Frisk90}. The strength of the shell effect at the Fermi
surface changes strongly in the prolate side while it stays almost
constant in the oblate side as a function of the magnitude of
deformation.
As the classical periodic orbit responsible for this asymmetry
between prolate and oblate shapes,
Frisk payed attention to a triangular orbit in the meridian plane whose
period changes little in the oblate side owing to the volume conservation
condition.

Arita used a power-law potential $\propto r^{\alpha}$~\cite{Ari12},
which can interpolate continuously between
the square-well potential and the harmonic oscillator potentials,
for the same kind of analysis
and obtained essentially the same conclusion as Frisk~\cite{Ari12}.
Concerning the classical periodic orbits responsible for the prolate-shape dominance,
Arita insisted the importance of bridge-orbit bifurcations.
He is also planning to treat the spin-orbit potential in the periodic orbit theory~\cite{Ari04}.

Hamamoto and Mottelson presented a plain explanation to the fact that
the level density in the Fermi level changes little
in the oblate side~\cite{HM09}.
In the Nilsson diagram of single-particle spectrum,
the degenerated levels belonging to a subshell at spherical shape
fans out (spreads) versus increasing magnitude of deformation in both prolate and oblate sides.
However, this fanning out is small for levels with low $\Lambda$ (the projection of orbital angular
momentum onto the symmetry axis) in the oblate side
because they are pushed down by many orbitals with the same quantum number
belonging to upper major shells.
Since the same spectrum leads to the same value of the shell effect,
no energy can be gained by deforming into oblate shapes.

However, Arita~\cite{Ari12} found a counterexample to this explanation.
For a low-power potential $\propto r^{1.1}$, the fanning is suppressed for low-$\Lambda$ orbitals
in {\em prolate} shapes, while he found no {\em oblate}-shape dominance.
Therefore, the explanation in terms of fanning applies indeed to nuclear-like potentials
but probably not to different types of potentials.
We think that the reason for this limitation is that
the explanation in terms of the fanning mechanism does not explicitly consider
the volume conservation condition which played an essential role in
Frisk's argument~\cite{Frisk90}.


Finally, we mention two other factors which have minor effects
to prefer prolate shapes.

One is the Coulomb interaction, which tends to elongate the nuclear shape rather than
flatten it in order to diminish positive electrostatic potential energy between protons.
However, this effect does not seem to play a decisive role to give
rise to the prolate-shape dominance because it increases as $Z^2$ and
is strongest in heavy nuclei while the actual dominance is already very
clear in middle-weight nuclei.
This effect will be revisited in our future paper for more detailed studies.

The other is the angular-momentum projection of mean-field solutions
into zero or low-spin spin states, which
has a tendency to lower the energy of prolate states relative to that
of oblate ones,
because the former have generally larger moments of inertia than
the latter.
This effect seems to have already been argued by Zickendraht~\cite{Zic85}
as the difference of the volume element for the collective coordinates
between prolate and oblate shapes, which is originated in the difference of the
available configuration space in the spherical shell model.
However, mean-field models and their phenomenological interactions
are determined to reproduce the experiments without angular momentum projection
and thus it is widely believed that
they includes the effects of the projection.
A quantitative examination of this belief will be a subject of our future study.

In Sec.~\ref{sec:methods}, we explain about the potentials,
our theoretical framework, some practical information of our calculations,
and definitions of the ratio of prolate nuclei.
In Sec.~\ref{sec:results}, we give the results and discussions on them.
In Sec.~\ref{sec:summary}, we summarize the conclusions of this paper.

\section{METHODS} \label{sec:methods}

We use the Woods-Saxon-Strutinsky~\cite{BDJ72} and the
Nilsson-Strutinsky methods~\cite{NTS69} 
to calculate the nuclear energy surface in the
plane spanned by the axially symmetric quadrupole and hexadecapole
deformation parameters and to determine the shape of the ground state
of even-even nuclei from Oxygen to superheavy elements between
drip lines.  For both methods, the single-particle Hamiltonian is a sum
of a free kinetic energy and a potential energy,
\begin{equation}
H = \frac{\bm{p}^2}{2m} + V,
\end{equation}
where $m$ is the bare nucleon mass (the average of a proton and a neutron masses).
The Nilsson and the WS potentials used for $V$ are described in
Secs.~\ref{sec:Nilsson_potential} and \ref{sec:WS_potential}, respectively.
The latter potential has a continuum part in the spectrum, which
causes difficulties in the standard version of the Strutinsky method.
Remedies we have chosen~\cite{TST10} are summarized shortly in
Sec.~\ref{sec:Strutinsky_method}.
Precision and other practical aspects of numerical computations are
described in Secs.~\ref{sec:precision} and \ref{sec:calculation_setup}.
The quantity which we study intensively in this paper is introduced in
Sec.~\ref{sec:ratiodefinition}.

\subsection{The Nilsson potential} \label{sec:Nilsson_potential}

The potential of the Nilsson model
\footnote{
 Note that the expression of the Nilsson potential
 in Ref.~\cite{TOST11} has a few misprints, which are now corrected.
}
is expressed as
\begin{equation}
\begin{split}
\hspace*{-0.3cm}
V(\bm{r})\!\!&=\!\!\frac{m}{2}(\omega^2_{\perp}x^2\!+\!\omega^2_{\perp}y^2\!+\!\omega^2_z z^2)
\!+\! \hbar\stackrel{\circ}{\omega_0} r_t^2\sqrt{\frac{4\pi}{9}}\epsilon_4 Y_{40}(\hat{\bm{r}}_t)\\
&\!\!-\!\!
2f_{ls}\kappa_{N}\hbar\stackrel{\circ}{\omega_0}\bm{l}_t \cdot \bm{s}
\!-\!f_{ll}\kappa_{N}\mu_{N}\hbar\stackrel{\circ}{\omega_0}(\bm{l}_t^2\!\!-\!\!\langle \bm{l}_t^2\rangle_N),
\end{split}
\end{equation}
where $\omega_\perp$ and $\omega_z$ are related to a quadrupole
deformation parameter $\epsilon_2$ through
$\omega_\perp \!\!=
\stackrel{\circ}{\omega_0}\!\!\left( 1\!+\!\frac{1}{3}\epsilon_2 \right)$
and
$\omega_z\!\!=
\stackrel{\circ}{\omega_0}\!\!\left(1\!-\!\frac{2}{3}\epsilon_2 \right)$
with $\stackrel{\circ}{\omega_0}$
determined by the condition of a volume conservation
$\omega_\perp^2\omega_z =\ \stackrel{\circ\,\,3}{\omega_0}$.
Denoted by $\epsilon_4$ is a hexadecapole deformation parameter.
Orbital and spin angular momenta are denoted as $\bm{l}$ and $\bm{s}$, respectively
and the subscript $t$ means the usage of the stretched coordinates.
The third term is a spin-orbit potential,
while the fourth term is called the $l^2$ potential.
The latter term proportional to the square of the orbital angular momentum
is used to simulate roughly
the change of the radial profile of the central potential
from $\propto r^2$ of the harmonic oscillator.
With a standard negative strength of the fourth term ($-f_{ll} \kappa_{N} \mu_{N}$)
and without the spin-orbit potential,
the spectrum is modified from that of the harmonic oscillator
toward that of the square-well potential.

We adopt the standard values given in Table~1 of Ref.~\cite{BR85}
for the parameters $\kappa_N$ and $\mu_N$
which are dependent on the harmonic oscillator shell $\nosc$.
By changing the set of values of the multipliers $(f_{ls}, f_{ll})$
from the standard values $(1,1)$,
one can examine how the combination of these two potentials
affects the prolate-shape dominance.
For example,
$f_{ls}$=0~(1) means no (standard) spin-orbit potential,
while $f_{ll}$=0~(1) corresponds to harmonic-oscillator (standard)
radial profile of the central potential.

In this paper, we calculate the ratio of prolate nuclei
for $31 \times 16$ = 496 kinds of artificially modified Nilsson potentials defined by
the combination of values of the multipliers $(f_{ls}, f_{ll})$
in ranges $-1.5 \leq f_{ls} \leq 1.5$ and $0 \leq f_{ll} \leq 1.5$
with $\Delta f_{ls}=\Delta f_{ll}=0.1$.

As a footnote,
the $\epsilon_{\lambda}$-parametrization of nuclear deformation
in the Nilsson potential is different
from the conventional $\beta_{\lambda}$-parametrization
based on the expansion of nuclear radius
in terms of the spherical harmonics~\cite{BDN89}.
Including only the $\lambda=2$ deformation and
requiring the ratio of axes is the same
between the two kinds of parametrizations, as an example,
one obtains
$\epsilon_2 \cong 0.95\,\beta_2$ for small deformations while
$\epsilon_2 \cong 1.14\,\beta_2$ for the ratio 2:1 of the ideal superdeformation.

In this paper, we describe quantities related to shapes in term of $\beta_2$.
For example, we give the thresholds to distinguish prolate or oblate
shapes from the spherical shape in terms of the values of $\beta_2$.
For the solutions of the Nilsson-Strutinsky method, we apply
these thresholds directory to $\epsilon_2$, not calculating the value of
$\beta_2$ corresponding to the value of $\epsilon_2$, for the sake of
simplicity.

\subsection{The Woods-Saxon potential} \label{sec:WS_potential}

The Woods-Saxon (WS) potential is
a finite-depth potential having a flat central and a steep surface parts.
It is no doubt a better approximation to the central part of the nuclear potential
$V\rs{CE}$ than the Nilsson potential.
The spin-orbit potential $V\rs{SO}$ is also reasonably expressed
using the gradient of this form of a potential.
The Coulomb potential $V\rs{CO}$ should be added for protons.
Then, the sum of these three potentials, which we also call the WS potential,
is expressed as
\begin{equation}
V = V\rs{CE}+V\rs{SO}+\frac{1}{2}(1-\tau_3)V\rs{CO},
\end{equation}
where $\tau_3$ is the third component of
the nucleon's isospin multiplied by 2 (1 for neutrons and $-1$ for protons),
\begin{equation}
V\rs{CE} = V\rs{WS}(\bm{r};V\rs{0CE},\kappa\rs{CE}, R\rs{0CE}, f_a a\rs{CE},\bm{\beta}),
\end{equation}
\begin{equation} \label{eq:VSO}
\begin{split}
V\rs{SO} &=
f_{ls} \lambda\rs{SO}\!\left(\frac{\hbar}{2m_{\rm red}c} \right)^2 \times \\
& [\nabla\!V\rs{WS}(\bm{r}; V\rs{0CE}, \kappa\rs{SO}, R\rs{0SO}, f_a a\rs{SO},\bm{\beta})] \cdot\\
& \left(\bm{\sigma}\!\times\frac{1}{i}\nabla\right).
\end{split}
\end{equation}
In Eq.~(\ref{eq:VSO}),
$m_{\rm red}=\frac{A-1}{A}m$ with $m$ being the bare nucleon mass is the reduced mass,
and $\bm{\sigma}$ is the Pauli matrix for the nucleon's spin.
The function $V\rs{WS}$ is defined by
\begin{equation}
\begin{split}
& V\rs{WS}(\bm{r}; V_{0}, \kappa, R_0, a, \bm{\beta}) \\
&=-V_0\left[ 1\pm \kappa\frac{N-Z}{A}\right]\frac{1}{1+\exp[{\rm{dist}}_\Sigma(\bm{r},\bm{\beta})/a]}.
\end{split}
\end{equation}
where dist$_\Sigma(\bm{r}, \bm{\beta})$ is the perpendicular
distance between the point $\bm{r}$ and the surface $\Sigma$
(taken with the minus sign inside the nucleus).

For axially symmetric nuclear shapes, the surface $\Sigma$ is defined by
\begin{equation}
R(\theta; R_{0}, \bm{\beta}) = R_{0}c_{v}(\bm{\beta})\left[
1+{\sum_{\lambda}}\beta_{\lambda}Y_{\lambda 0}(\theta)\right],
\end{equation}
where $\bm{\beta} \equiv \{\beta_{\lambda}\}$ and
$c_{v}(\bm{\beta}) $ is a normalization factor to conserve the volume.

As the standard parameter set which corresponds to the actual nuclear potential,
we employ the universal parameter set given in Table~1 of \'{C}wiok et
al.~\cite{CDN87} for the main results.
Precisely speaking, we do {\it not} use the depth of the central potential as it is, but modify it for each nucleus in a way explained in Sec.~\ref{sec:Strutinsky_method} and Ref.~\cite{TSS02}.
We also use five other parameter sets, which we call
Wahlborn~\cite{BW60},
Rost~\cite{Ro68},
Chepurnov~\cite{Ch67},
Wyss-1~\cite{BVC10},
and Wyss-2~\cite{WysPrivate05} (see Table I of Ref.~\cite{SS09}),
to confirm the independence of the conclusions from the choice of the
standard parameter set.

Similarly to the Nilsson case,
two constants $f_{a}$ and $f_{ls}$ are introduced by which
to multiply the standard values of the surface diffuseness
$a$ and the spin-orbit potential strength $\lambda\rs{SO}$, respectively.
By changing the combination of the values of these multipliers $(f_{a}, f_{ls})$
from the standard values $(1,1)$,
we examine how the prolate-shape dominance depends on the combined effects of the
surface diffuseness and the spin-orbit potential.

It should be noted that the centrifugal potential enhances the effect
of diffuseness on orbits having large angular momentum by pushing
their wave functions out strongly on the surface.  Thus, the effect of
changing $a$ by several tens of percents is quite large despite the
fact that the standard value $a=0.7~{\rm fm}$ is much smaller than the
nuclear radius $\sim 10^{1}~{\rm fm}$.

The multiplier $f_{ls}$ of the WS potential
is approximately equivalent to $f_{ls}$ of the Nilsson potential
because both of them are the multipliers to the spin-orbit potential.
Indeed, as we will show in Sec.~\ref{sec:results}, the same value of $f_{ls}$
seems to correspond to the same situation between the two kind of potentials
as far as the ratio of prolate-shape nuclei concerns.

On the other hand, $f_{a}$ of the WS potential
and $f_{ll}$ of the Nilsson potential work in different ways
to change (literally or effectively) the radial profile of the potential.
A finite-height cavity (i.e., square-well) potential is obtained in the limit
$f_{a} \rightarrow 0$ of the WS potential and it can also be
approximated by the Nilsson potential with a certain value
of $f_{ll}$ ($>1$, dependent on the mass number $A$).
A harmonic oscillator potential ($\propto r^2$) is obtained
by setting $f_{ll}=0$ of the Nilsson potential and it can also
be approximately expressed by a certain value
of $f_{a}$ ($>1$, depending on $A$) for the WS potential.
We will show in Sec.~\ref{sec:mainresult}
that $f_{a}=0$, $1$, and $2$
correspond very roughly to
$f_{ll}=1.5$, $1$, and $0$,
respectively,
as far as the ratio of prolate-shape nuclei concerns,
which is an average over Oxygen to superheavy elements.
One may also think roughly that
WS potentials with $f_{a} \gtrsim 2$ cannot be expressed by Nilsson potentials
while Nilsson potentials with $f_{ll} \gtrsim 1.5$ or $f_{ll}<0$ cannot be
expressed by WS potentials.

In this paper, we calculate the ratio of prolate nuclei
for $31 \times 19$ = 589 kinds of artificially modified WS potentials defined by
the combination of values of the multipliers $(f_{ls}, f_{a})$
in ranges $-1.5 \leq f_{ls} \leq 1.5$ and $0.2 \leq f_{a} \leq 2$
with $\Delta f_{ls}=\Delta f_{a}=0.1$.


\begin{figure}[!tbh] 
\includegraphics[width=.48\textwidth]{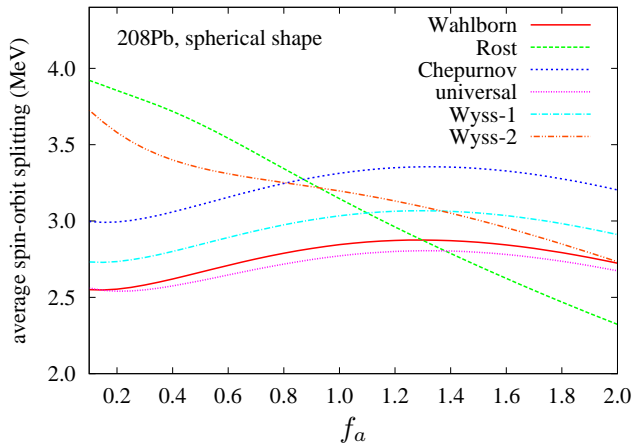}
\vspace*{-5mm}
\caption{
  Average spin-orbit splitting in units of MeV of $^{208}$Pb
  versus a multiplier $f_{a}$ to the surface diffuseness
  for six sets of parameters of the WS potential.
}
\label{fig:lssall}
\end{figure} 

Finally, we remark on the dependence of spin-orbit splittings,
$\epsilon_{j=l-\frac{1}{2}} - \epsilon_{j=l+\frac{1}{2}}$,
on the surface diffuseness.
Concerning the Nilsson potential,
the effects of two multipliers $f_{ls}$ and $f_{ll}$ are orthogonal
and $f_{ll}$ never affects spin-orbit splittings.
As for the WS potential, however,
spin-orbit splittings can be affected by $f_{a}$
through the radial form factor $\nabla V_{\rm WS}$
in $V_{\rm SO}$ of Eq.~(\ref{eq:VSO}).
(All the WS parameter sets postulate $a\rs{SO} = a\rs{CE}$.)
However, the net effect turns out to be small on the average
after the integral is done over the radius.

In Fig.~\ref{fig:lssall}, we show
the average value of spin-orbit splittings in $^{208}$Pb
as a function of $f_{a}$.
The potential is spherical and the parameter set is the universal set.
In calculating the average, we consider all the pairs of levels
whose energies are negative throughout the plotted region
$0.1 \leq f_{a} \leq 2$.
The number of states in each pair of levels ($4l+2$) is used as the
weight in the averaging procedure.
We do not distinguish proton and neutron levels for this figure.
For the Wahlborn, Chepurnov, universal, and Wyss-1 parameter sets,
the variation in the average splitting is small ($\sim 10\%$).
For the Rost and Wyss-2 parameter sets,
the variation is larger (monotonically decreasing).
We find that these different behaviors are originated in the difference
between $R\rs{SO}$ and $R\rs{CE}$, i.e.,
the splitting does not change very much if $R\rs{SO}=R\rs{CE}$,
while it is a monotonically decreasing (increasing) function of $f_{a}$ 
if $R\rs{SO} < R\rs{CE}$ ($R\rs{SO} > R\rs{CE}$).
The spin-orbit splitting of each pair of levels (not shown) behaves in the same
way as the average (shown) except when the upper level is close to zero
energy.

\begin{figure}[!tbh] 
\includegraphics[width=.48\textwidth]{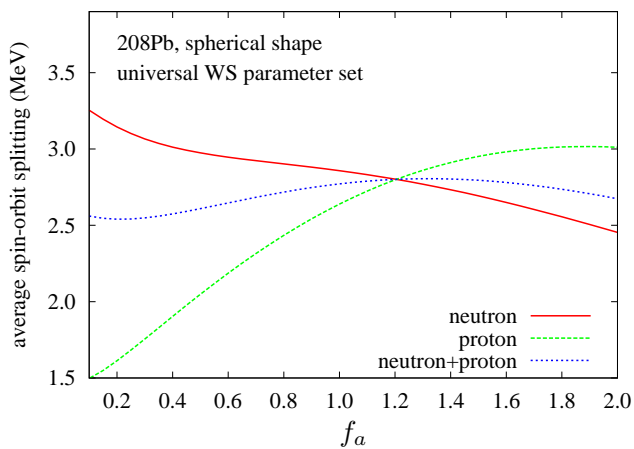}
\vspace*{-5mm}
\caption{
  Spin-orbit splittings in units of MeV of $^{208}$Pb averaged over neutron levels,
  proton levels, and all the levels versus $f_{a}$
  calculated with the universal parameter set for the WS potential.
}
\label{fig:lssuniversal}
\end{figure} 

The universal parameter set is exceptional in the sense that
$R\rs{SO}/R\rs{CE} < 1$ for neutron and $> 1$ for proton.
We show the average over neutron levels and that over proton levels separately
for the splittings calculated with this parameter set
in Fig.~\ref{fig:lssuniversal}.
The curve labeled as ``neutron+proton'' is the same as the curve
for this parameter set in Fig.~\ref{fig:lssall}.
One can see that the neutron splitting decreases by 25\%
while the proton splitting increases by 100\% in the
plotted interval of $f_{a}$.
It is not clear whether
the effects cancel between neutrons' decreasing and
and protons' increasing splittings concerning the quantities like the ratio of prolate nuclei.
From this point of view, the universal parameter set may not be
the most suitable set for the purpose of this paper.
However, it is not a major demerit since the principal results of this paper
like Fig.~\ref{fig:rp1} and Fig.~\ref{fig:rp1otherwsps} do not look
very different between any parameter sets.
Anyway, we regard the universal parameter set as the standard
and mainly show the results calculated with that parameter set.

\subsection{The Strutinsky method and its extensions} \label{sec:Strutinsky_method}

The shape and energy of the ground state of each nucleus
are determined by the shell correction method~\cite{Strt}.
However, the conventional method cannot be applied to finite-depth potentials
owing to the continuum part of the spectrum~\cite{NWD94}.
In order to cope with this difficulty,
we employ an improved shell correction method which we have recently proposed
(the reference density method~\cite{TST10}) based on the Kruppa prescription~\cite{Kruppa98}.

In shell-correction approaches,
the total energy of a nucleus is assumed to be expressed as a sum of
the macroscopic and microscopic parts,
\begin{equation}
E=E_{\rm mac}+E_{\rm mic} \equiv E_{\rm LDM}+(E_{\rm BCS}-\tilde{E}_{\rm BCS}),
\end{equation}
where $E_{\rm mac}=E_{\rm LDM}$ is the energy of the liquid-drop
model, whose parameters are taken from Ref.~\cite{Af99}.
The energies, $E_{\rm BCS}$ and $\tilde{E}_{\rm BCS}$, are
calculated as
\begin{equation}
\left(
\begin{array}{c}
E_{\rm BCS} \\
\tilde{E}_{\rm BCS}
\end{array}
\right)
= \int_{-\infty}^{+\infty}
\left(
\begin{array}{c}
v^2(\epsilon)g(\epsilon)\\
\tilde{v}^2(\epsilon)\tilde{g}(\epsilon)
\end{array}
\right)
\epsilon d\epsilon \\
-\frac{1}{G}
\left(
\begin{array}{c}
\Delta^2 \\
\tilde{\Delta}^2
\end{array}
\right),
\end{equation}
where the occupation probability $v^2(\epsilon)$ and the pairing gap $\Delta$
($\tilde{v}^2(\epsilon)$ and $\tilde{\Delta}$) are
obtained by solving the BCS equation for the seniority-type pairing
interaction, whose matrix elements are a constant $G$,
with discrete level density $g(\epsilon)$
(Strutinsky smoothed level density $\tilde{g}(\epsilon)$).

The interaction strength $G$ is determined
such that the smoothed pairing gap agrees with the empirical smooth trend,
\begin{equation} \label{eq:average_pairing_gap}
\tilde{\Delta}=\fctd \times 13\,A^{-1/2}  (\rm{MeV}),
\end{equation}
where the multiplier $\fctd$ is one (which corresponds to actual nuclei)
for most of the calculations except in Sec.~\ref{sec:pairdep}
where we try three other values ( $\fctd=0$, $0.7$, $1.2$ ).


The depth of the central potential $V\rs{CE}$ is adjusted for the
consistency between the microscopic (in the Thomas-Fermi approximation)
and macroscopic Fermi energies for spherical shapes neglecting the
spin-orbit potential.  This adjustment is necessary in order to treat
nuclei far from the $\beta$ stability line because the available
parameter sets of the WS potential do not reproduce the location of
the drip lines predicted by the macroscopic part of energy.
This adjustment is also indispensable when
$f_{a} \not= 1$, because the Fermi level is strongly affected by $f_{a}$.

More details on the methods of calculations are described in Ref.~\cite{TST10}.

\subsection{Numerical precision} \label{sec:precision}

We have confirmed that
numerically calculated single-particle spectrum of the WS potentials
is sufficiently precise to calculate the ratio of prolate nuclei.
It should be noted that usually recommended size and frequency
of the oscillator basis are not guaranteed to be sufficient
to diagonalize Hamiltonians when potentials are artificially modified,
i.e., when $f_{ls}$ or $f_{a}$ is not one.

We prepare the basis of diagonalization as eigenfunctions of a Hamiltonian of
a nucleon of mass $m$ in a harmonic oscillator potential,
whose frequency is determined in such a way that
the major shell spacing is equal to
$\facomega \hbar \omega = \facomega \times 41 A^{-1/3}$ MeV.
Here, $\facomega$ is an adjustment factor, usually takes on 1.2.
When $f_{a} > 1$, the spatial extension becomes larger so that
the oscillator length of the basis $\propto \facomega^{-1/2}$
should be increased, i.e.,
$\facomega$ should be decreased.
We determine empirically $\facomega$ as a function of $f_{a}$,
\begin{equation} \label{eq:facoemga}
\facomega =
  \left\{
    \begin{array}{ll}
      1.2         & (0 < f_a \leq 0.5), \\
      1.4-0.4 f_a & (0.5 < f_a \leq 1.5), \\
      0.8         & (f_a \geq 1.5),
    \end{array}
  \right.
\end{equation}
so as to obtain more precise results with smaller basis.

The single-particle basis is truncated according to the maximum number
of the oscillator quanta $\nosc$.  
In this paper, we diagonalize mostly in a
subspace $\nosc \leq 20$.
Concerning the calculations for 
Figs.~\ref{fig:rp1}, \ref{fig:rp2}, \ref{fig:mass_rp}, and \ref{fig:rp1otherwsps},
we consider only $\nosc \leq 16$ because of the required large computations.
As we show in Sec.~\ref{sec:mainresult},
the results are quite close to each other between the calculations
with $\nosc \leq 16$ and $\nosc \leq 24$ as far as the ratio of
prolate nuclei concerns.

\subsection{Set up of the numerical calculations} \label{sec:calculation_setup}

For each potential specified by a set of multipliers
$(f_{ls}, f_{ll})$ or $(f_{ls}, f_{a})$,
we calculate the ground-state shape
for all the even-even nuclei
with $8\leq Z\leq 126$ and $8\leq N \leq 184$
between the proton and neutron drip lines
predicted by the Bethe-Weiz\"acker (macroscopic) mass formula (2148 nuclei).
To find the ground-state shape for each of these 2148 nuclei with
each of the 589 kinds of the WS potentials,
we calculate the nuclear total energy surface
versus the quadrupole and hexadecapole deformation
parameters $(\beta_2, \beta_4)$
at $51 \times 21 = 1071$ points in ranges
$-0.5 \leq \beta_2 \leq 0.5$ with $\Delta \beta_2=0.02$
and
$-0.3 \leq \beta_4 \leq 0.3$ with $\Delta \beta_4=0.03$.
For the Nilsson-Strutinsky calculations of the 496 kinds of
the Nilsson potentials, $\beta_\lambda$ should be read as $\epsilon_\lambda$.

The scale of required numerical computation is quite large.
In order to obtain each of the six panels
Fig.~\ref{fig:rp1}~(a) and Fig.~\ref{fig:rp1otherwsps} (a--e)
corresponding to the different parameter sets of the WS potentials,
we have to calculate the total energy
589 $\times$ 2148 $\times$ 1071 = $1.4 \times 10^9$ times,
each time of which is for a nucleus using a given-shape potential.
We have performed such computations in 64 threads in a PC cluster having
eight Intel Core i7-920 CPU (2.66GHz, four physical = eight logical cores in a CPU).
It has typically taken 28 hours for a physical core to complete a nuclear chart for a given
combination of ($f_{ls}, f_{a}$), and 20 days for the cluster to complete all the combinations,
when we treat the WS potential in an oscillator space truncated by $\nosc \leq 16$.
Compared with the WS potentials,
it takes negligibly small time to perform the corresponding
Nilsson-Strutinsky calculations.

\subsection{Definition of the ratio of prolate nuclei} \label{sec:ratiodefinition}

After completing all the Strutinsky-method calculations,
we count the number of prolate, oblate, and spherical nuclei in each nuclear chart
to calculate the ratio of prolate nuclei $\rp$ as a measure of the prolate-shape dominance.
We use two different definitions of the ratio in order to demonstrate
that the main conclusions of this paper are not affected by the choice of the definition.

By $\numnuclo$, $\numnucls$, and $\numnuclp$,
let us denote the number of nuclei whose ground states are
oblate ($\beta_2 \leq -0.05$),
spherical ($|\beta_2| < 0.05$),
and prolate ($\beta_2 \geq 0.05$),
respectively,
among the 2148 even-even nuclei defined in Sec.~\ref{sec:calculation_setup},
with those having positive Fermi levels excluded.
Using these numbers, the ratio of prolate nuclei
may be defined as
\begin{equation} \label{eq:rp2}
\rpp = \frac{\numnuclp}{\numnuclp+\numnuclo}
\end{equation}

In Ref.~\cite{TS01}, however, we have proposed a different definition
in order to exclude shape transitional nuclei which often have
an almost flat bottom in the nuclear energy curve ranging typically between
$-0.2 \lesssim \beta_2 \lesssim 0.2$.
It is better to exclude them
because precise ground states of such nuclei are
superpositions of prolate and oblate shape states
owing to large zero-point quantum fluctuations.
There is little point whether the minimum happens to be located in the prolate or oblate side.
After an extensive examination of the landscapes of total energy curves of
all the even-even nuclei,
we found that excluding spherical and transitional nuclei
is nearly equivalent to including only those nuclei which have
both prolate and oblate minima (discarding very shallow ones~\cite{TS01}).
Hence we have defined alternatively the ratio of prolate nuclei as
\begin{equation} \label{eq:rp1}
\rp = \frac{\numnuclwdp}{\numnuclwdp+\numnuclwdo}
\end{equation}
where $\numnuclwdp$ ( $\numnuclwdo$ ) denotes
the number of nuclei which has both oblate and prolate minima
and the prolate (oblate) minima has a lower energy.
(Superscript ``DM'' stands for ``double-minimum'').

In Sec.~\ref{sec:mainresult}, we will demonstrate that
the main conclusions of this paper is not altered essentially
whether we use either $\rp$ or $\rpp$.
The advantage of $\rp$ is the exclusion of transitional nuclei
while the merit of $\rpp$ is only simplicity.
Hence we mainly use $\rp$ in this paper.

\section{RESULTS} \label{sec:results}

\subsection{Ratio of prolate nuclei versus potential parameters} \label{sec:mainresult}

\begin{figure}[!ht] 
\includegraphics[width=.48\textwidth]{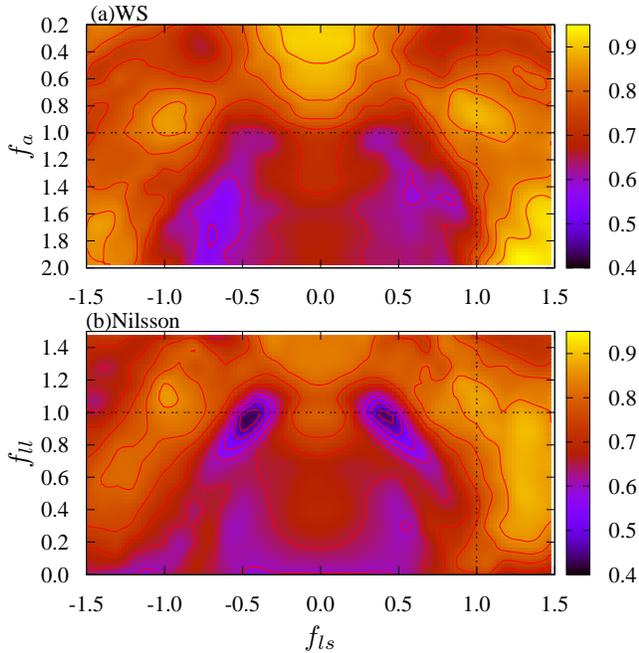}
\vspace*{-5mm}
\caption{\label{fig:rp1}
  The ratio of prolate nuclei $\rp$ for the WS (a) and the Nilsson (b) potentials.
  The abscissa is the multiplier $f_{ls}$ to the spin-orbit potential for both panels.
  The ordinate is the multiplier $f_{a}$ to the surface diffuseness $a$ of the WS potential (a)
  and the multiplier $f_{ll}$ to the $\bm{l}^2$ potential of the Nilsson potential (b).
  The universal parameter set is used as the standard one for the WS potential.
}
\end{figure} 

In Fig.~\ref{fig:rp1}, the ratio of prolate nuclei $\rp$ defined by Eq.~(\ref{eq:rp1})
is shown as a function of $f_{ls}$ and $f_{a}$ for the WS potential (a)
and $f_{ll}$ and $f_{ls}$ for the Nilsson potential (b).
We use the universal parameter set~\cite{CDN87} for the WS potential.
Fourth order Legendre polynomials in $f_{ls}$ and those in $f_a$ (or $f_{ll}$)
are used for the interpolations to draw the contours.

\begin{figure}[tb] 
\includegraphics[width=.48\textwidth]{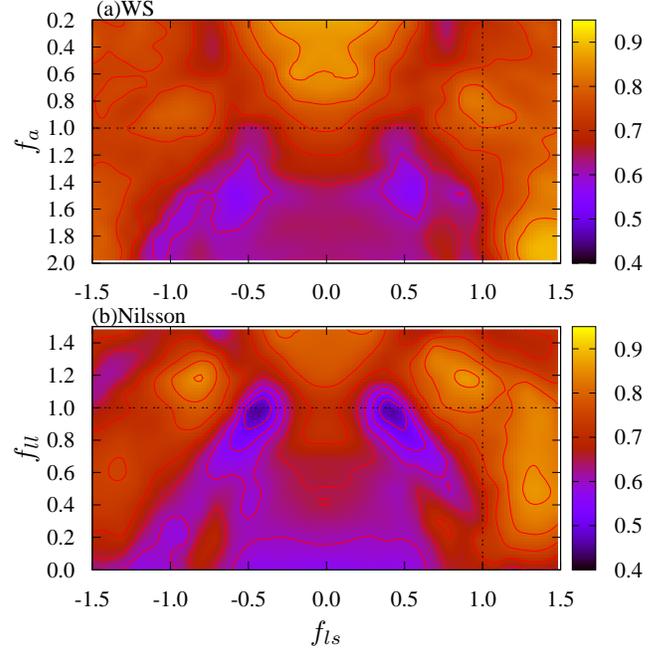}
\vspace*{-5mm}
\caption{
  Same as Fig.~\ref{fig:rp1}
  but for a different definition of the ratio of prolate nuclei $\rpp$.
}
\label{fig:rp2}
\end{figure} 

\begin{figure}[!ht] 
\includegraphics[width=.48\textwidth]{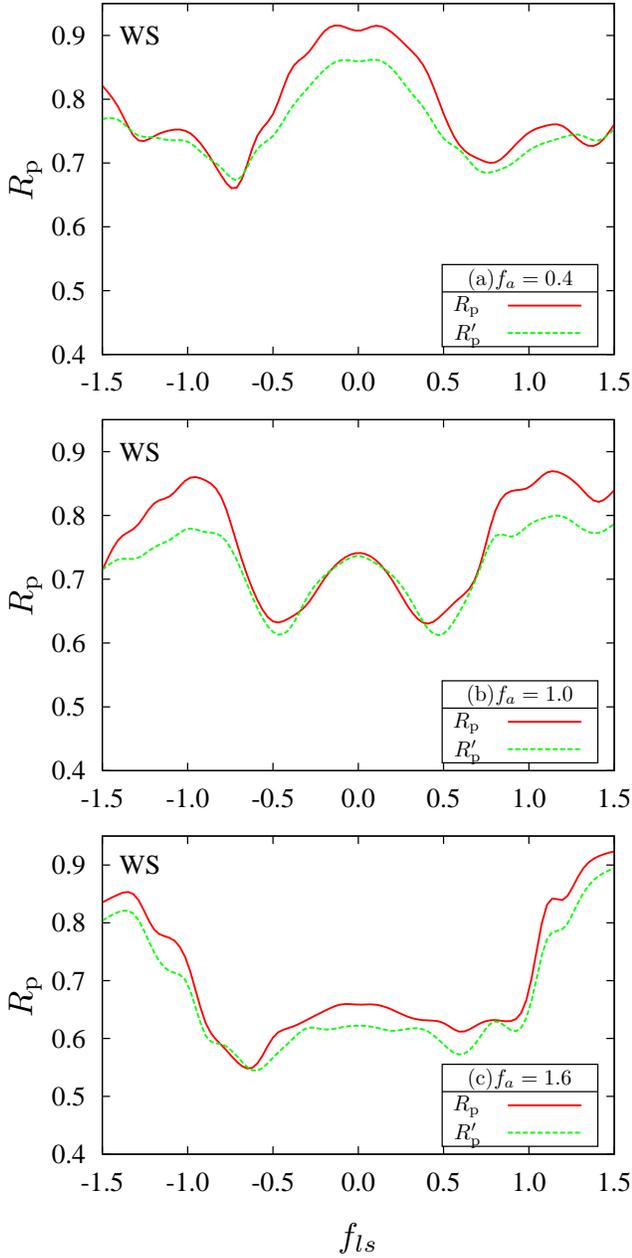}
\vspace*{-5mm}
\caption{The ratio of prolate nuclei $\rp$ and $\rpp$ for the WS
  potential for (a) $f_a=0.4$, (b) $f_a=1$, and (c) $f_a=1.6$
  versus the multiplier to the spin-orbit strength $f_{ls}$.
}
\label{fig:rp1fa}
\end{figure}

\begin{figure}[!ht] 
\includegraphics[width=.48\textwidth]{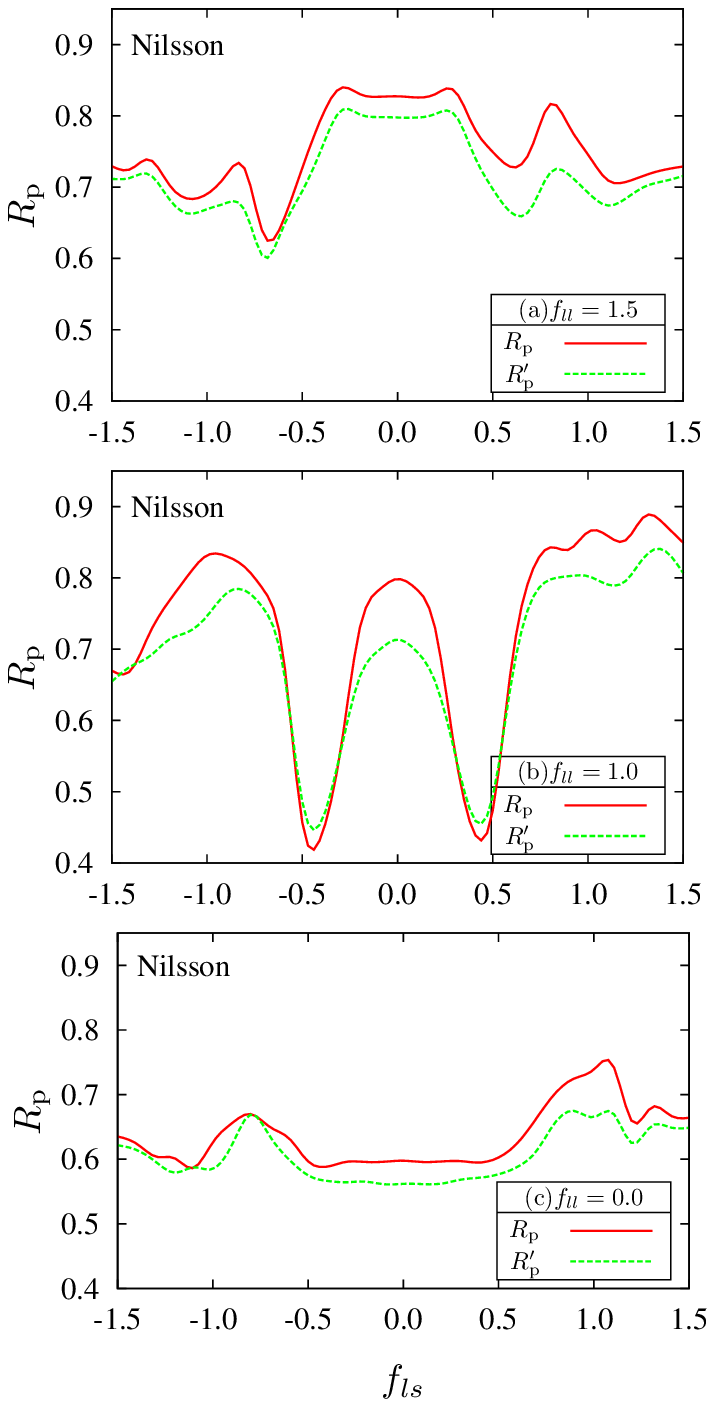}
\vspace*{-5mm}
\caption{The ratio of prolate nuclei $\rp$ and $\rpp$ for the Nilsson
  potential for 
(a) $f_{ll}=1.5$, (b) $f_{ll}=1$, and (c) $f_{ll}=0$
  versus the multiplier to the spin-orbit strength $f_{ls}$.
}
\label{fig:rp1fll}
\end{figure}

\begin{figure}[!ht] 
\includegraphics[width=.48\textwidth]{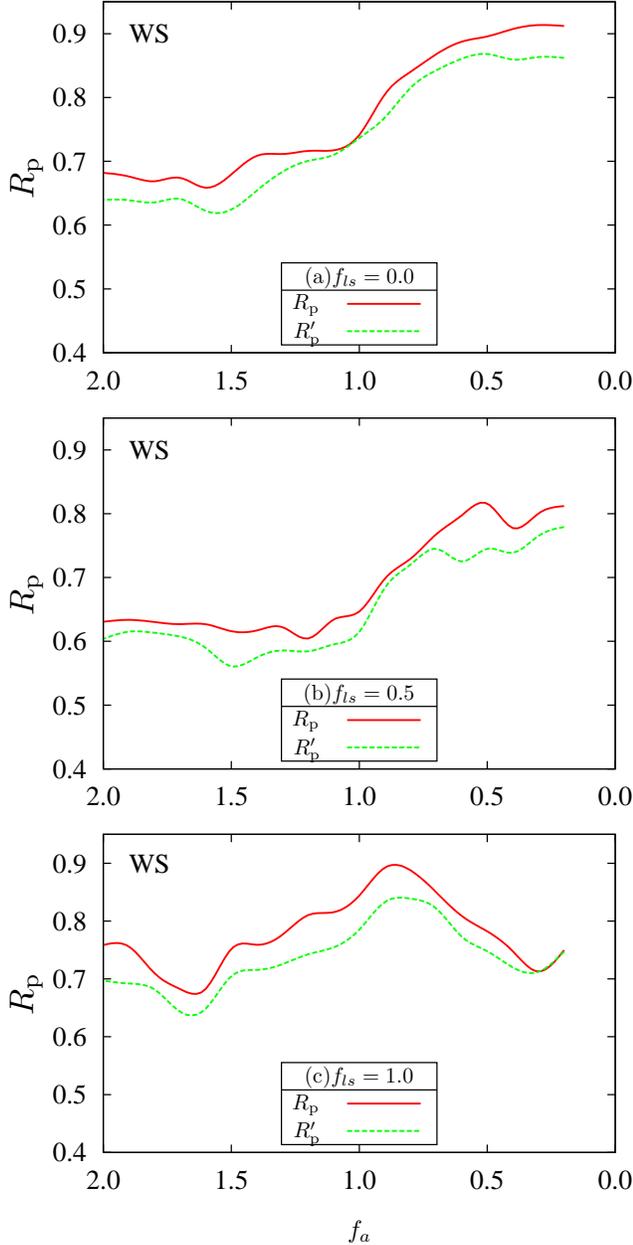}
\vspace*{-5mm}
\caption{The ratio of prolate nuclei $\rp$ and $\rpp$ for the WS
  potential for (a) $f_{ls}=0$, (b) $f_{ls}=0.5$, and (c) $f_{ls}=1$
  versus the multiplier to the surface diffuseness $f_{a}$.
}
\label{fig:rp1fl}
\end{figure}

\begin{figure}[!ht] 
\includegraphics[width=.48\textwidth]{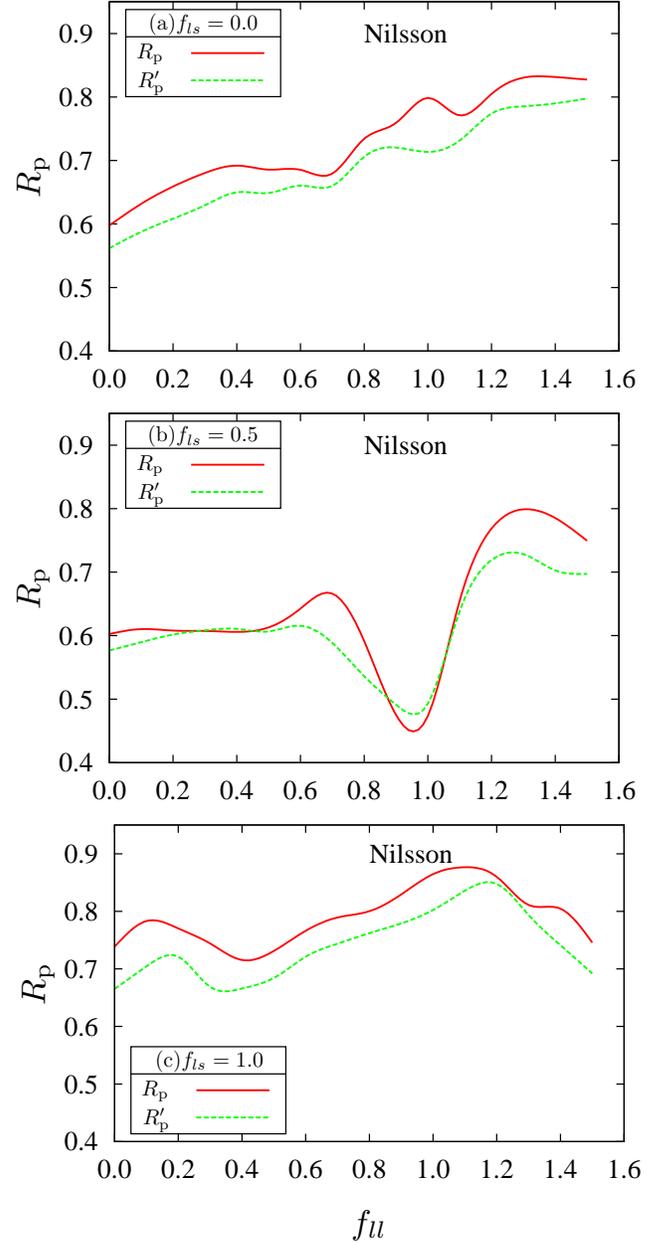}
\vspace*{-5mm}
\caption{The ratio of prolate nuclei $\rp$ and $\rpp$ for the Nilsson
  potential for (a) $f_{ls}=0$, (b) $f_{ls}=0.5$, and (c) $f_{ls}=1$
  versus the multiplier to the $\bm{l}^2$ term $f_{ll}$.
}
\label{fig:rp1fls}
\end{figure}

In Fig.~\ref{fig:rp2}, the ratio of prolate nuclei $\rpp$ of a simpler definition
of Eq.~(\ref{eq:rp2})
is shown in the same manner as $\rp$ in Fig.~\ref{fig:rp1}.
One can see the behavior of $\rpp$ is quite similar to that of $\rp$.
It can also be confirmed in detail in Figs.~\ref{fig:rp1fa}--\ref{fig:rp1fls}, which
show curves for $\rp$ and $\rpp$ in various cross sections of the panels
in Figs.~\ref{fig:rp1} and~\ref{fig:rp2}.
The distance between the two curves is mostly $\sim$ 0.05.
The largest discrepancy is 0.11
at $(f_{ls},f_{ll})=(0,1)$ in Fig.~\ref{fig:rp1fll}(b).
We have also checked that a third definition
\begin{equation} \label{eq:rppp}
  \rppp = \frac{\numnuclp}{\numnuclp+\numnucls+\numnuclo}
\end{equation}
does not change the behavior very much, either (no figures).
Therefore, the following discussions are independent of the definition of
the ratio of prolate nuclei.

Result for the Nilsson potential shown in
Fig.~\ref{fig:rp1}(b) is essentially the same
as Fig.~1 of Ref.~\cite{TS01}, except that
the area of $f_{ll} < 0$ is not shown in this paper.

In a line $f_{ls}=0$ (Fig.~\ref{fig:rp1fls}(a)),
i.e., when there is no spin-orbit potential,
$\rp$ ($\rpp$) increases monotonically as a function of $f_{ll}$,
from 60\% (56\%) at $f_{ll}=0$ (i.e., the harmonic oscillator potential)
to 80\% (71\%) at $f_{ll}=1$ (corresponding to the actual surface thickness).
This behavior agrees with Frisk's argument~\cite{Frisk90} that
a cavity(-like) potential prefers prolate shapes more than oblate ones.

In a line $f_{ll}=1$ (Fig.~\ref{fig:rp1fll}(b)),
i.e., with the surface diffuseness of actual nuclei,
$\rp$ as well as $\rpp$ oscillates strongly as a function of $f_{ls}$.
It takes local maximum values at $f_{ls} \simeq \pm 1$
and local minimum values at $f_{ls} \simeq \pm 0.5$.
It means a strong interference between the effects of the $ls$ and $\bm{l}^2$ potentials.
Minima of $\rp$ are located close to this line $f_{ll}=1$ 
where $\rp \simeq 40\%$, i.e., oblate shapes dominate over prolate shapes
if the spin-orbit potential is weakened by 50\% while keeping the
surface diffuseness of the potentials at the standard value.

The result for the WS potential is shown in
Fig.~\ref{fig:rp1}(a).
\footnote{
Note that 
we plotted $\rpp$ by mistake in Fig.~1 of Ref.~\cite{TOST11}, 
which is consequently the same figure as Fig.~\ref{fig:rp2}(a), not Fig.~\ref{fig:rp1}(a).
}
It looks quite similar to 
Fig.~\ref{fig:rp1}(b) for the Nilsson potential.

In a line $f_{ls}=0$ (Fig.~\ref{fig:rp1fl}(a)),
$\rp$ ($\rpp$) increases almost monotonically as a function of $-f_{a}$,
from 68\%  (64\%) at $f_{a}=2$
  to 91\%  (86\%) at $f_{a}=0.2$.
This behavior during the transition
from the harmonic oscillator to a cavity-like potentials
agrees qualitatively with that of the Nilsson potential and
with Frisk's argument~\cite{Frisk90}.
Quantitatively, however,
$\rp=\rpp=73\%$ at $f_{a}=1$ (actual nuclear central potential)
is not sufficiently dominant.
The situation is worse in a line $f_{ls}=0.5$ (Fig.~\ref{fig:rp1fl}(b)).
Only by the assist of the actual-strength spin-orbit potential, i.e., $f_{ls}=1$
(Fig.~\ref{fig:rp1fl}(c)), $\rp$ can exceed 80\% when $f_{a}=1$.

As for the Nilsson potential,
this assist of the spin-orbit potential works stronger
to change the landscape of the curve.
When $f_{ls}=0.5$, it generates a large-amplitude oscillation versus $f_{ll}$
(Fig.~\ref{fig:rp1fls}(b)), whose minimum at $f_{ll}=0.95$  
corresponds to the minimum at 
$(f_{ls}, f_{ll})=(0.45, 1)$ in Fig.~\ref{fig:rp1fll}(b).
It is this oscillation that has moved two local minima upward
in two-parameter plots of $\rp$ and of $\rpp$
compared with those for the WS potential:
For the WS potential (Fig.~\ref{fig:rp1}~(a)), the minima are
located around a line $f_a \simeq 1.6$, not $f_a = 1$, while
for the Nilsson potential (Fig.~\ref{fig:rp1}(b)), they are
just in a line $f_{ll}=1$.
The amplitude of the oscillation in a line $f_{ll}=1$ (Fig.~\ref{fig:rp1fll}(b))
is larger than
that in a line $f_{a}=1.6$ (Fig.~\ref{fig:rp1fa}(c))
or
that in a line $f_{a}=1$ (Fig.~\ref{fig:rp1fa}(b))
by a factor $\simeq 3$.
The reason for the exaggeration of the amplitude of the oscillation by the Nilsson potential
may be the approximate treatment of the radial profile of the potential
in terms of the $l^2$ term
and the affinity between this $l^2$ term and the $ls$ term in a sense
that both of them contain the same orbital angular momentum operator $\bm{l}$.
Anyway, we have confirmed a fact that
the interference also exists for the WS potential.

\begin{figure}[!ht] 
\includegraphics[width=.48\textwidth]{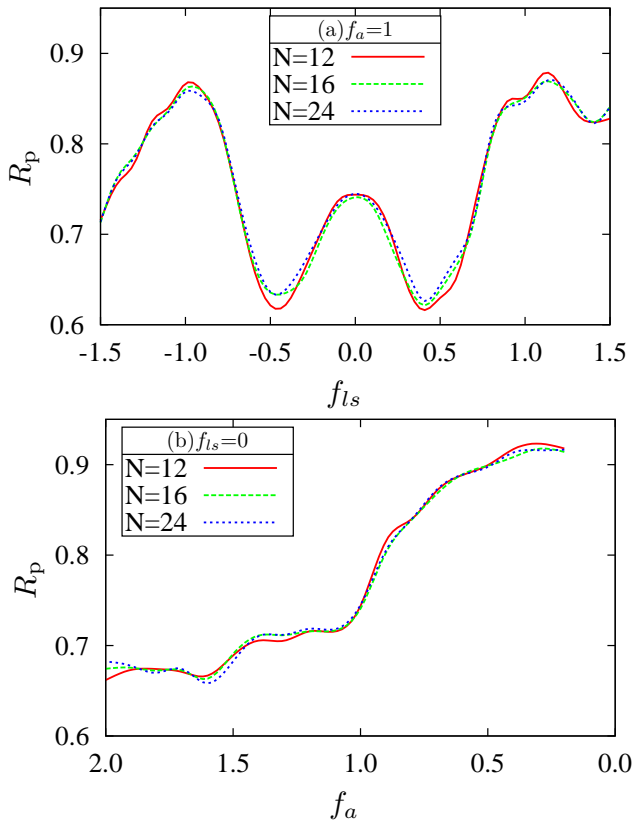}
\vspace*{-5mm}
\caption{Dependence of the ratio of prolate nuclei $\rp$ on the size of the oscillator basis
specified by $\nosc$, which is the maximum oscillator quantum number
of the single-particle basis.
In panel (a), $\rp$ is plotted versus the multiplier to the spin-orbit potential $f_{ls}$
while the surface diffuseness is frozen at the standard value ($f_{a}=1$).
In panel (b), $\rp$ is plotted versus the multiplier to the surface diffuseness $f_{a}$
while the spin-orbit potential is kept turned off ($f_{ls}=0$).
}
\label{fig:nmaxdep}
\end{figure} 

In Fig.~\ref{fig:nmaxdep}, we show that our single-particle basis is sufficiently large.
Panel (a) is the same as Fig.~\ref{fig:rp1fa}(b)
and panel (b) is the same as Fig.~\ref{fig:rp1fl}(a)
except that results with three different sizes of basis are compared.
One can see in both panels that
the curve for $\nosc \leq 16$ is quite close to the curve for $\nosc \leq 24$.

\subsection{The effect of the pairing strength} \label{sec:pairdep}

\begin{figure}[!ht] 
\includegraphics[width=.48\textwidth]{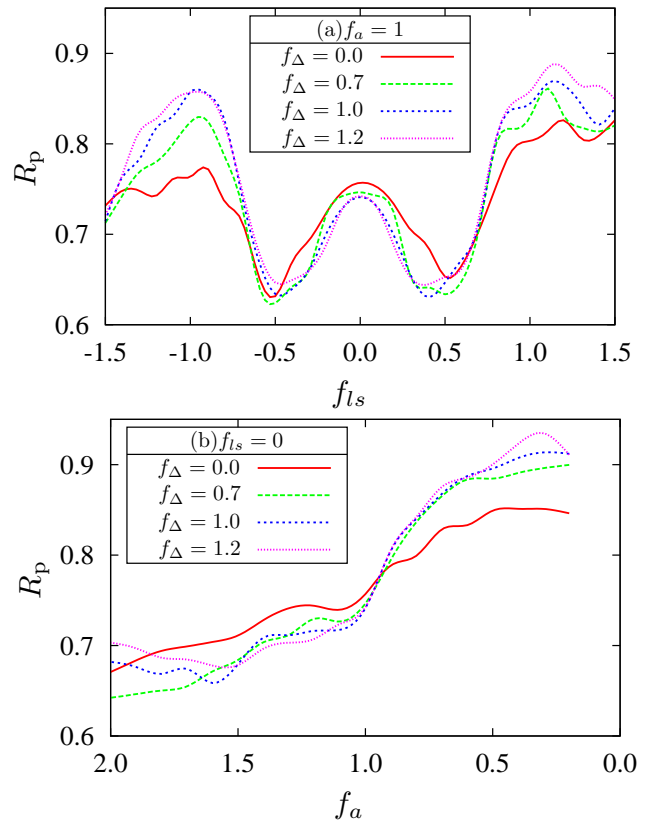}
\vspace*{-5mm}
\caption{
  The dependence of the ratio of prolate nuclei $\rp$ on the pairing strength
  for the WS potential.
  Four kinds of curves correspond to different values of a modification factor $\fctd$
  which determines the strength of the paring force. For $\fctd=1$, the empirical
  trend is used. For $\fctd=0$, there is no pairing correlation.
  Panel (a): as a function of the spin-orbit strength $f_{ls}$ with a normal
  diffuseness $f_{a}=1$.
  Panel (b): as a function of the diffuseness $f_{a}$ without a spin-orbit potential.
}
\label{fig:pairdep}
\end{figure} 

In Fig.~\ref{fig:pairdep}, the ratio of prolate nuclei $\rp$ is shown for four
values of the multiplier $\fctd$ appearing in Eq.~(\ref{eq:average_pairing_gap})
for the WS potential.
The pairing strength can be controlled with this multiplier.

In panel (a), $\rp$ is shown as a function of $f_{ls}$
while the surface diffuseness is the standard one ($f_{a}=1$).
The short dash curve labeled as $\fctd=1.0$ is the same as
the solid curve in Fig.~\ref{fig:rp1fa}(b).
One can see that a stronger pairing increases the amplitude of oscillation.
The same tendency has also been found for the Nilsson potential~\cite{TSS02}.
The change is largest at prolate-shape dominant points of $f_{ls} \sim \pm 1$,
where $\rp$ is increased further from a high value to a higher value.
This can be understood that generally shallower minima in oblate nuclei are
more easily dissolved and merged with the other minima
to form a single spherical minimum than generally deeper
minima in prolate nuclei.

In panel (b), $\rp$ is drawn versus $f_{a}$
while the spin-orbit potential is kept turned off ($f_{ls}=0$).
The short dash curve labeled as $\fctd=1.0$ is the same as
the solid curve in Fig.~\ref{fig:rp1fl}(a).
One can see an overall trend
that stronger pairing correlation increases the magnitude of the slope.
Pairing correlation increases $\rp$ at $f_{a} \lesssim 1$,
which may be understood by the same interpretation given in the last paragraph.
Pairing correlation decreases $\rp$ at $1 \lesssim f_{a} \lesssim 1.5$.
At $f_{a} \gtrsim 1.5$, curves are reordered and
the trend of monotonic increase of $\rp$ versus $-f_{a}$ is reversed
for $\fctd \geq 1$.
These behaviors require a different interpretation yet to be explored.

\subsection{Mass dependence of the ratio of prolate nuclei} \label{sec:massdep}

\begin{figure}[!ht] 
\includegraphics[width=.48\textwidth]{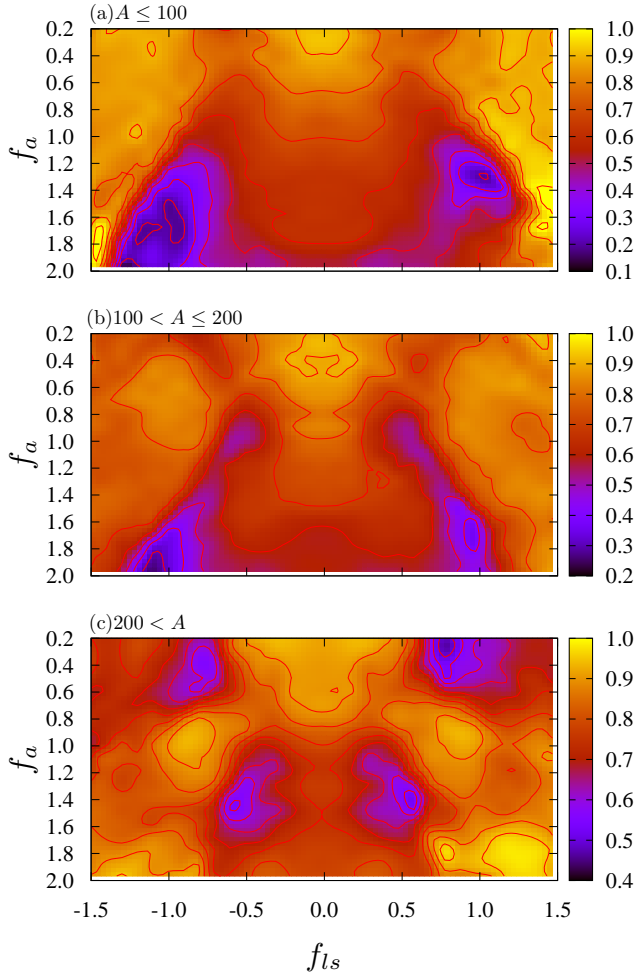}
\vspace*{-5mm}
\caption{Same as in Fig.~\ref{fig:rp1}(a) but decomposed into
  a light $A\leq 100$~(a),  a medium $100 < A \leq 200$~(b),
  and a heavy $200 < A$~(c) mass regions.
}
\label{fig:mass_rp}
\end{figure} 

We decompose Fig.~\ref{fig:rp1}(a) into three mass regions,
$A \leq 100$, $100 < A \leq 200$, and $A > 200$ to obtain
the three panels of Fig.~\ref{fig:mass_rp}.
One can see that the patterns in these three mass regions
are rather different; the numbers of oscillations in both directions
($f_a$ and $f_{ls}$) are larger in the heavier mass region.
Namely, the ratio of prolate nuclei changes more rapidly
when the potential is modified.
This pattern change can be understood by the fact that the level density
is larger in heavier system, where the prolate- and oblate-driving
orbits coexist in a narrow energy range at the Fermi surface
and a modification of the potential more easily changes
the equilibrium deformation.
Alternatively, one can think that the pattern is ``shrunken''
in two axes from the light to the heavy regions,
although new pronounced low $\rp$ areas appear in the range of small
$f_a \lesssim 0.5$ in the heavy mass region (Fig.~\ref{fig:mass_rp}(c)).
The shrinkage in the $f_a$ axis may be ascribed to
the ratio of a constant thickness $a$ to the nuclear radius $\propto A^{1/3}$,
while that in the $f_{ls}$ axis may be related to the
ratio of the spin-orbit splitting $\propto \nosc \propto A^{1/3}$ to
the major shell spacing $\propto A^{-1/3}$.
This shrinkage may cause a blurring of the pattern averaged over all the mass regions.
It may be possible that, by employing a properly $A$-scaled
parametrizations for the surface diffuseness $a$ and
the spin-orbit potential strength $\lambda\rs{SO}$,
instead of simple constant multipliers $f_{a}$ and $f_{ls}$,
one obtains a little sharper pattern than Fig.~\ref{fig:rp1}.

\subsection{Absolute number of nuclei of prolate, oblate, and spherical shapes}
\label{sec:absnum}

\begin{figure}[!ht] 
\includegraphics[width=.48\textwidth]{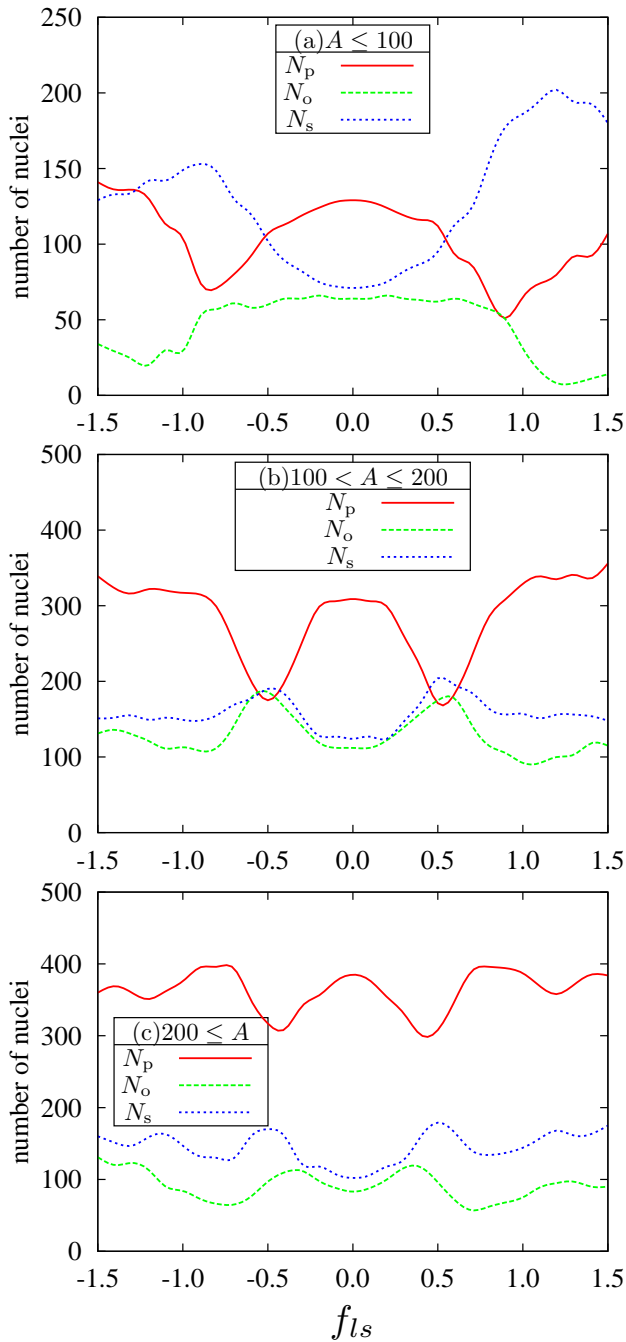}
\vspace*{-5mm}
\caption{
The number of prolate ($\numnuclp$), oblate ($\numnuclo$), and spherical nuclei ($\numnucls$)
in a light (a), a medium (b), and a heavy (c) mass regions
as a function of $f_{ls}$ (with $f_{a} = 1$).
}
\label{fig:mass_num}
\end{figure} 

In Fig.~\ref{fig:mass_num}, we show not ratios but absolute numbers
of nuclei of prolate, oblate, and spherical shapes at their ground states,
i.e., $\numnuclp$,  $\numnuclo$, and $\numnucls$ defined in Sec.~\ref{sec:ratiodefinition}.
The numbers of nuclei are plotted versus $f_{ls}$ while $f_{a}$ is frozen at 1.
Because these numbers depend on the mass number more strongly than ratios,
we discuss separately the light, medium, and heavy mass regions
introduced in Sec.~\ref{sec:massdep}.
As it is discussed in the previous subsection,
one observes considerable differences in three mass regions;
the three numbers, $\numnuclp$,  $\numnuclo$, and $\numnucls$,
oscillate more frequently in the heavy mass region.
In Fig.~\ref{fig:mass_num}(a), which shows the numbers of nuclei in the light mass region,
the oscillation in $\numnuclp$ for $|f_{ls}| \leq 0.8$ can be ascribed to
shape transitions of the ground state of each nuclei
between prolate and spherical shapes
because $\numnuclo$ is almost constant in that interval of $f_{ls}$.
In Fig.~\ref{fig:mass_num}(b), which is for the medium mass region,
the amplitude of oscillation in $\numnuclp$ is most pronounced among the three mass regions.
The oscillation is attributed evenly to
shape transitions between prolate and spherical shapes
and those between prolate and oblate shapes.
In Fig.~\ref{fig:mass_num}(c), which corresponds to the heavy mass region,
the amplitude of the oscillation in $\numnuclp$ is reduced.
This may be a consequence of the fact that,
from the light to the heavy regions,
the prolate nuclei continues to become more and more dominant over
spherical and oblate nuclei for the normal diffuseness $f_a=1$.

\subsection{Map of deformation on the nuclear chart}
\label{sec:deformationmap}

\begin{figure}[htb] 
\includegraphics[width=.40\textwidth]{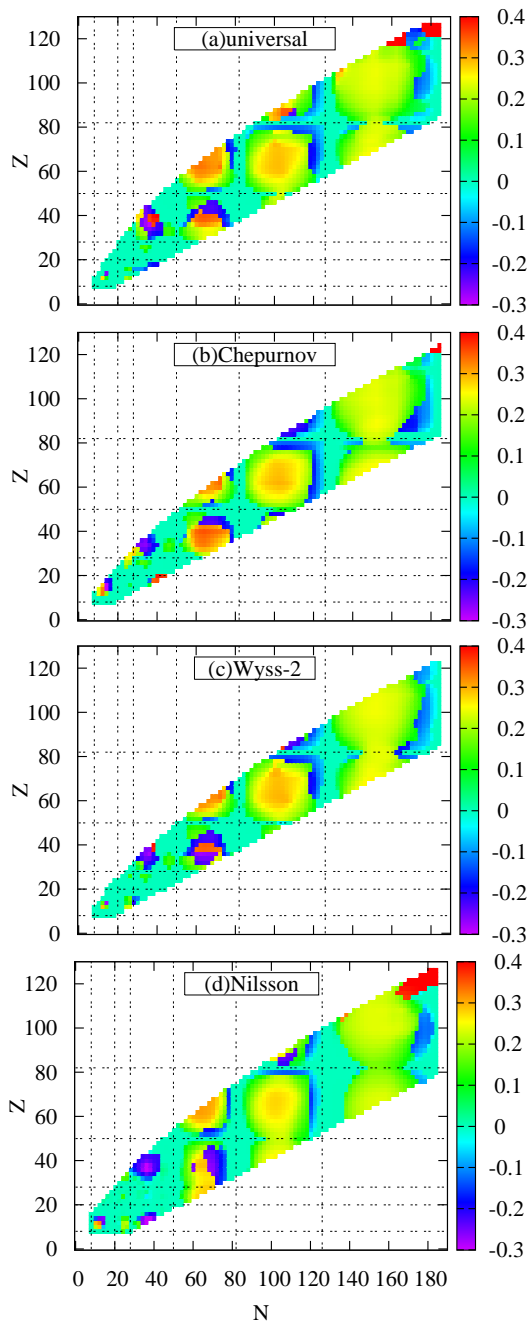}
\vspace*{-5mm}
\caption{Quadrupole deformation parameter $\beta_2$ of the ground states
  of even-even nuclei
  calculated with standard parameters for the WS potential (a--c) and
  the Nilsson potential (d).
  The used parameter set for the WS potential are
  the universal (a), the Chepurnov (b), and the Wyss-2 (c).
}
\label{fig:defmapstd}
\end{figure}

In Fig.~\ref{fig:defmapstd}, we plot the quadrupole deformation parameter
$\beta_2$ of the ground states of even-even nuclei
in the ($N$, $Z$) plane.
In panels (a--c), the WS potential is used.
The parameter sets are
the universal parameter set which is the choice of this paper (a),
the Chepurnov parameter set~\cite{Ch67} which is one of the classical (b),
and the Wyss-2~\cite{WysPrivate05} parameter set which is one of the latest (c).
In panel (d), the Nilsson potential is used and $\epsilon_2$ is plotted instead of $\beta_2$.
Comparing the predictions of these potential parameter sets,
one sees that the locations of prolate and oblate nuclei are roughly the same
among the four panels. This means that
the conclusions of this paper is not restricted to the mainly employed
universal parameter set.

With a more careful look at these maps, one finds
that a rectangular area $28 < Z \lesssim 40$, $50 < N < 82$ is the area
where the largest number of nuclei have differently predicted shapes by
different parameter sets; the prediction of the Nilsson potential
is also very different from those of the WS potentials.
In this area, nuclei of large prolate deformations and those of large oblate deformations
are located side-by-side directly, meaning the first order phase transition in the shape.
A similar situation that the distinct prolate and oblate minima coexist
in the same energy region has been extensively discussed
in the $A\sim 80$ region, see e.g. Ref.~\cite{NDB85}.
Compared to the $A\sim 80$ region, this rectangular area is much wider
so that it may be a more suitable region to study the issue of
the shape coexistence and/or the shape phase transition
in the light of recent developments of unstable beam facilities.
It should be noticed that a smaller area $28 < Z,N \lesssim 40$
has a similar characteristic, which provides typical examples
of the shape coexistence phenomenon~\cite{TTO96},
where the spherical shape also play an important role.

\begin{figure}[htb] 
\includegraphics[width=.40\textwidth]{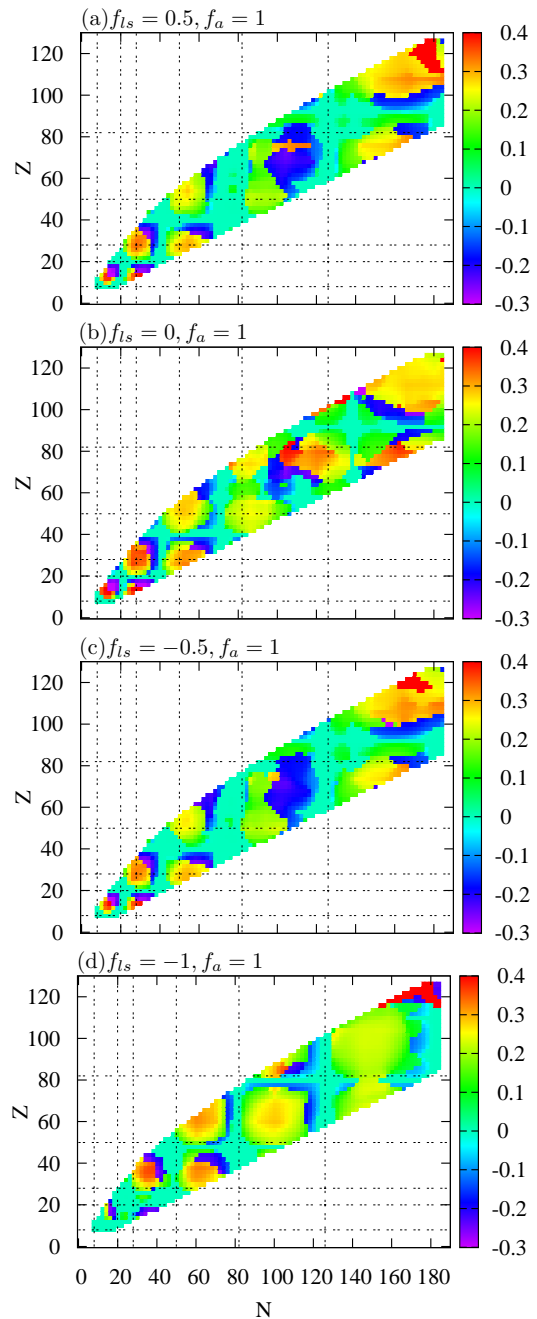}
\vspace*{-5mm}
\caption{Quadrupole deformation parameter $\beta_2$ of the ground states
  of even-even nuclei
  using the universal parameter set for the WS potential
  with multipliers $f_{a}=1$ and
  $f_{ls}=0.5$ (a),
  $0$ (b),
  $-0.5$ (c), and
  $-1$ (d).
  Horizontal and vertical dot lines indicate the magic numbers
  of actual nuclei, not of the modified potentials.
}
\label{fig:defmaposc}
\end{figure} 

The deformation map changes drastically when the multipliers are changed.
In Fig.~\ref{fig:defmaposc},
we plot the quadrupole deformation $\beta_2$
predicted with the universal parameter set for the WS potential
modified with a multiplier $f_{ls}=0.5$ (a), $0$ (b) ,$-0.5$ (c), and $-1$ (d)
while $f_{a}$ is frozen at 1.
Comparing these four panels together with Fig.~\ref{fig:defmapstd}(a) for $f_{ls}=1$,
one sees a substantial change of the distribution of the shapes
caused by varying the value of $f_{ls}$.

One also notices that positive and negative values of $f_{ls}$ of the same
magnitude (i.e., Fig.~\ref{fig:defmapstd}(a) and Fig.~\ref{fig:defmaposc}(d),
Fig.~\ref{fig:defmaposc}(a) and Fig.~\ref{fig:defmaposc}(c))
lead to very similar distributions.

For $f_{ls}=\pm 1$,
oblate-shape nuclei are located in a narrow region just below major-shell closures.
On the other hand, for $f_{ls}=\pm 0.5$,
they occupy a large area in the latter half of a major-shell filling.
The largest of such areas are in $50 \leq Z \le 80$, $90 \leq N \le 120$
in Fig.~\ref{fig:defmaposc}(a,c).

It is interesting that
parameters $f_{ls}=0$ and $f_{ls}=\pm 1$, which lead to similar levels of prolate-shape dominance,
have completely different shape distribution maps.
This seems mainly because, when the spin-orbit potential is missing ($f_{ls}=0$),
subshell closures are strong enough to subdivide conventional major shells.

\subsection{Dependence on WS parameters}

\begin{figure}[htb] 
\includegraphics[width=.48\textwidth]{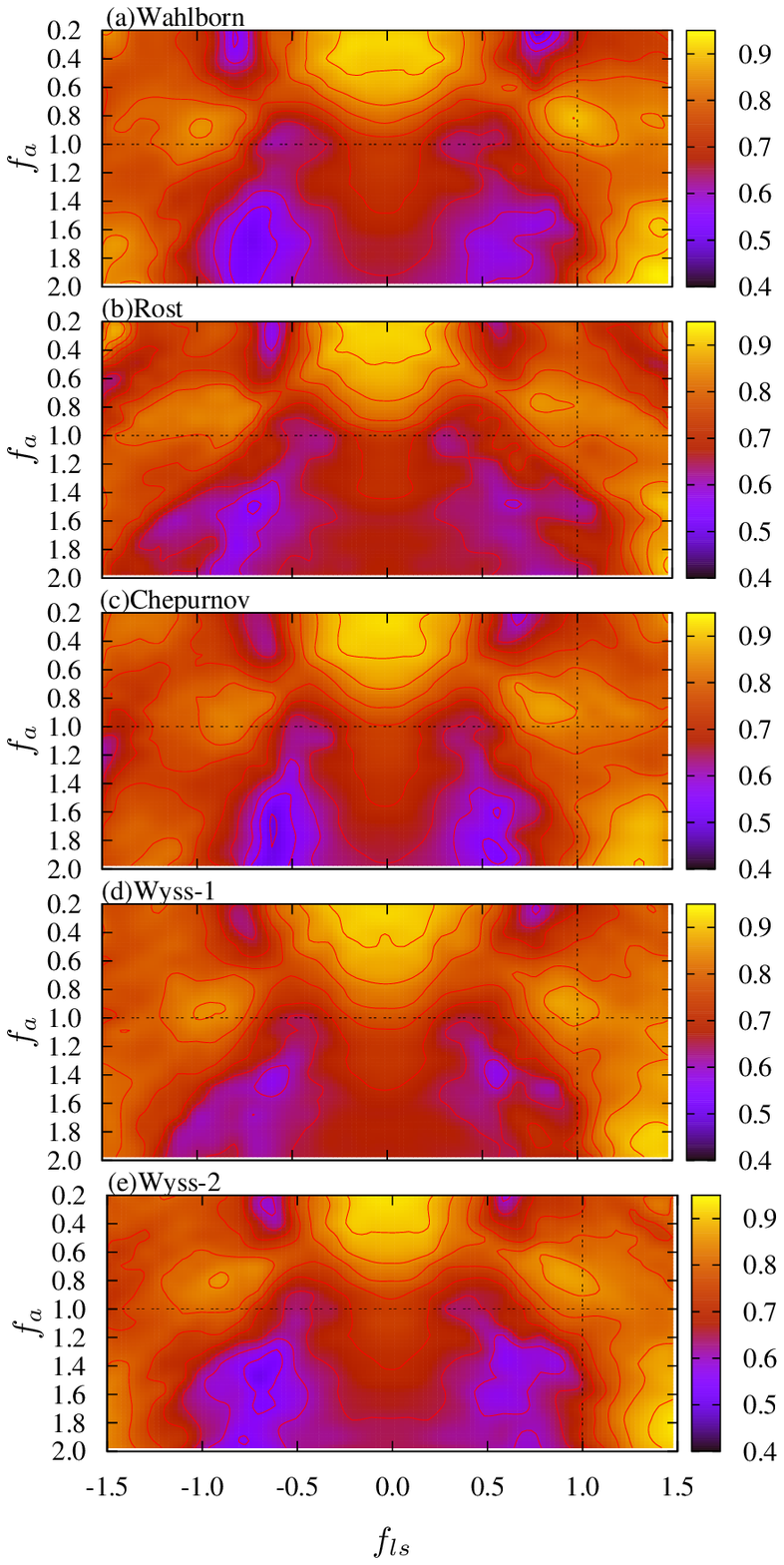}
\caption{Ratio of prolate nuclei $\rp$ calculated by using different WS parameter sets
as the standard,  the Wahlborn (a), the Rost (b), the Chepurnov (c),
Wyss-1 (d), and Wyss-2 (e).
}
\label{fig:rp1otherwsps}
\end{figure} 

In Fig.~\ref{fig:rp1otherwsps}, we show
the ratio of prolate nuclei $\rp$ for various WS parameter sets.
Combined with Fig.~\ref{fig:rp1}~(a), all the six sets of
the WS potentials are investigated.
In all the figures the basic pattern of the contour plot is the same,
although the precise positions of the low $\rp$ regions are slightly
shifted in each case.  Especially, the physical points, $f_{ls}=f_a=1$,
is near one of the highest points,
indicating the strong prolate-shape dominance for all parameter sets.
By comparing them, one can say that the conclusions of this paper
are not essentially altered by changing the standard potential.

\subsection{Single-particle level densities}


\begin{figure}[!ht] 
\includegraphics[width=.48\textwidth]{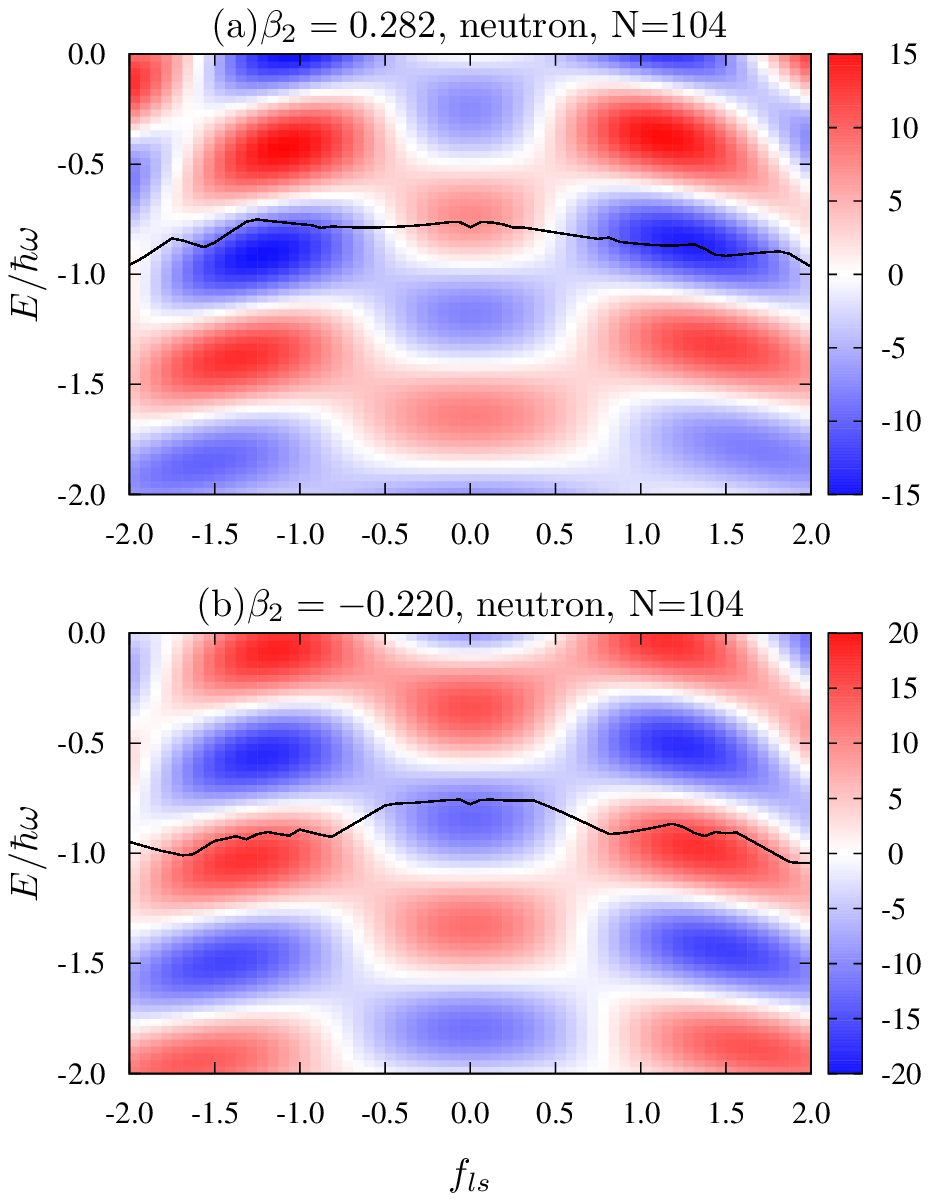}
\caption{
  Oscillating part of the neutron's single-particle level density
  $\delta \tilde{g}(\epsilon)$ of $^{172}\rm{Er}$
  as a function of the multiplier to the spin-orbit potential $f_{ls}$
  and the single-particle energy divided by $\hbar \omega$.
  The shape of the potential is prolate in panel (a) and oblate in panel (b).
  The Fermi level is designated with a black curve.
}
\label{fig:lvd}
\end{figure} 

We give an example to illustrate the way how the potential parameter changes the
shape of a nucleus through the change of the level density.
We choose a nucleus $^{172}_{\phantom{1}68}\rm{Er}_{104}$, which is located near the center of
a region $52\leq Z\leq 80, \ 90 \leq N \leq 110$, where the ground state is prolate
for the standard potential and oblate for a modified potential
with $(f_{ls},f_{a})=(0.5,1)$ as shown in Fig.~\ref{fig:defmaposc}(a).

In Fig.~\ref{fig:lvd}, the oscillating part of the neutron's level density
defined by,
\begin{equation}
  \delta \tilde{g}(\epsilon)=\tilde{g}_{\gamma=0.5}(\epsilon)-\tilde{g}_{\gamma=1.2}(\epsilon),
\label{eq:lvdg}
\end{equation}
of the ground state of this nucleus
is plotted as a function of the multiplier to the spin-orbit
potential $f_{ls}$ (abscissa) and the single particle energy in unit of $\hbar\omega$
(ordinate).
The Fermi level for neutrons is designated with a black curve.
The quantity $\gamma$ in Eq.~(\ref{eq:lvdg}) is the Strutinsky smoothing
parameter in unit of $\hbar\omega$, and the value $\gamma=0.5$ is chosen
to clearly show the oscillation due to the shell effect
relative to the average level density with $\gamma=1.2$,
with which all Strutinsky smoothed quantities are calculated.

Fig.~\ref{fig:lvd}(a) shows the level density at a prolate shape ($\beta_2=0.282$),
in which the level density at the Fermi level is low (high) at $f_{ls} = 1$ ($0.5$).
Fig.~\ref{fig:lvd}(b) shows the level density at an oblate shape ($\beta_2=-0.220$),
in which the level density at the Fermi level is high (low) at $f_{ls} = 1$ ($0.5$).
The change from a prolate to an oblate shape of this nucleus
due to the change of $f_{ls}$ from $1$ to $0.5$ can be explained in this way
in the macroscopic-microscopic theory.

\subsection{Prolate, oblate, and spherical ratios
as functions of Fermi energy }

\begin{figure}[!ht] 
\includegraphics[width=.48\textwidth]{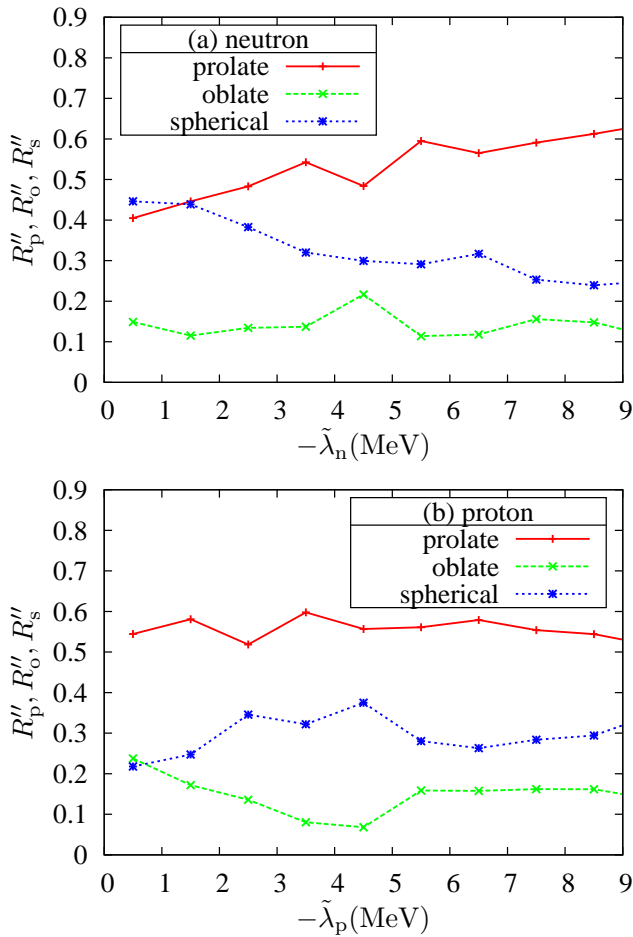}
\caption{
Ratios of prolate, oblate, and spherical nuclei as functions
of the Strutinsky smoothed Fermi energy $\tilde{\lambda}$
for the neutrons (a) or for the protons (b).
The parameter set is the standard ($f_{ls}=f_a=1$) universal set.
}
\label{fig:lmdratio}
\end{figure} 

As it is explained in detail in Sec.~\ref{sec:calculation_setup},
we have calculated the ground state masses of 2148 even-even nuclei
in the nuclear chart
for each modified set of parameters of the WS potential.
In view of the results,
there are many interesting findings
apart from the prolate-shape dominance.
Among them, we present the result on how the deformation of unstable nuclei
changes compared with the stable nuclei.
In Fig.~\ref{fig:lmdratio},
three ratios of the prolate, oblate, and spherical nuclei,
i.e., $\rppp$ defined by Eq.~(\ref{eq:rppp}) and
\begin{equation} \label{eq:ropp}
  \ropp = \frac{\numnuclo}{\numnuclp+\numnucls+\numnuclo}, \;\;\;
  \rspp = \frac{\numnucls}{\numnuclp+\numnucls+\numnuclo}
\end{equation}
are shown
as functions of the neutron (panel (a)) or proton (panel (b)) Fermi energy.
Note that these ratios are now defined with including $\numnucls$
in contrast to the definitions of $\rp$ and $\rpp$ used in previous subsections which exclude it.
For every one MeV interval of the calculated Fermi energy,
we count the numbers of nuclei $\numnuclp$, $\numnuclo$, and $\numnucls$,
calculate their ratios $\rppp$, $\ropp$, and $\rspp$,
and plot the ratios with the abscissa at the center of the interval in Fig.~\ref{fig:lmdratio}.
The Strutinsky smoothed Fermi energy is used as the calculated Fermi energy.
There are a few other choices for the Fermi energy;
for examples,
half of the two nucleon separation energy, $S_{\rm 2n}/2$ ($S_{\rm 2p}/2$),
or the chemical potential obtained by the BCS calculation.
The latter is not well-defined in the case of vanishing pairing gap.
The former choice has been checked that the resultant plot is very similar
to Fig.~\ref{fig:lmdratio}.

It has been sometimes discussed that in the weakly-bound system
the wave functions of the orbits near the Fermi surface spread out
and then the shell effects play a minor role.  Hence, the ratio
of spherical nuclei increases in weakly-bound unstable nuclei.
Such a trend is clearly seen in Fig.~\ref{fig:lmdratio}(a);
the ratio of prolate nuclei decreases from 0.6
at $\tilde{\lambda}_{\rm n}\approx -8.5$ to 0.4
at $\tilde{\lambda}_{\rm n}\approx -0.5$ in balance with the increase
of the spherical ratio.  Even the prolate-shape dominance
is inverted in the limit of weak-binding for neutrons,
$-\tilde{\lambda}_{\rm n} \rightarrow 0$.
As for protons, there is no such trend; the ratios are almost constants
as the Fermi energy is increased and the spherical ratio of nuclei
even slightly decreases.
This is because the existence of the Coulomb barrier prevents
the weak-binding situation even in the limit of vanishing Fermi energy.

\section{SUMMARY} \label{sec:summary}

We search
for the origin of the prolate-shape dominance of nuclear ground state
deformation in (the combinations of) the properties
of the single-particle potential for nucleons.
We employ the Woods-Saxon (WS) potential and apply to it
the macroscopic-microscopic theory modified for the treatment of the
continuum part of the spectrum.
We change the surface thickness, the strength of the spin-orbit potential,
and the strength of pairing correlations to examine their influences on the ratio of
prolate nuclei over more than two thousand even-even nuclei.
We observe strong interference between the effects of the surface
thickness and the spin-orbit potential, an especially interesting
consequence of which is an oscillation versus the spin-orbit potential
strength.
We also find that pairing correlations enhance prolate-shape dominance.

These results are compared with the results of the Nilsson potential
to elucidate some of the special features of the Nilsson potential, especially an
exaggeration of the amplitude of this oscillation
for the ratio of prolate nuclei.

We repeat calculations for six different parameter sets of the WS potential.
We find these parameter sets can be classified according to
the relation between the radii of the central and the spin-orbit potentials,
which determines the effect of the surface diffuseness on spin-orbit splittings.
However, as far as the behavior of the ratio of prolate nuclei concerns,
different parameter sets do not change the conclusions of this paper substantially.

The difference of the definition of the ratio of prolate nuclei is also shown to be
unimportant.
The error in the ratio of prolate nuclei due to the truncation of the
oscillator basis to express single-particle wavefunctions is shown to be sufficiently small.

Maps of ground state quadrupole deformation are shown and discussed in detail for
the physics behind the ratio of prolate nuclei, which is only a single number.
An example is given to illustrate the relation between the level density and
the shape of the ground state in the macroscopic-microscopic theory.
The ratios of prolate, oblate, and spherical nuclei are investigated also
as functions of the Fermi energy.  It is found that the spherical ratio
increases and even exceeds the prolate ratio in the limit
of weak-binding for the neutron rich unstable nuclei.

\section{ACKNOWLEDGMENTS}
The authors would like to thank Dr.~N.~Onishi and Dr.~K.~Arita for
discussions
and Dr. R.~Wyss for providing an unpublished parameter set ``Wyss-2''
for the Woods-Saxon potential.
This work was supported by Grant-in-Aid for Scientific Research (C)
No.~18540258 and No.~22540285 from Japan Society for the Promotion of Science.
A part of the formula manipulations were carried out on the computer system
at YITP in Kyoto University.
\vspace*{10mm}


\begin{thebibliography}{99}


\bibitem{BM75} 
  A. Bohr and B.R. Mottelson, {\em Nuclear Structure}
  (Benjamin, New York, 1975) Vol.\ 2.
\bibitem{Strt}
V.~M.~Strutisky, Sov. J. Nul. Phys. {\bf 3}, 449 (1966);
Nucl. Phys. A {\bf 95}, 420 (1967).

\bibitem{Frisk90}H.~Frisk, Nucl. Phys. {\bf A511}, 309 (1990).

\bibitem{Ari12}
  K.~Arita, arXiv:1202.5631.
\bibitem{TS01}N.~Tajima and N.~Suzuki, Phys. Rev. {\bf C64}, 037301 (2001).
\bibitem{NTS69} 
  S.G.~Nilsson, C.F.~Tsang, A.~Sobiczewski, Z.~Szyma\'{n}ski,
  S.~Wycech, C.~Gustafson, I.~Lamm, P.~M\"{o}ller and B.~Nilsson,
  Nucl.\ Phys.\ {\bf A131}, 1 (1969).

\bibitem{TOST11}
S.~Takahara, N.~Onishi, Y.~R.~Shimizu, and N.~Tajima,
  Phys. Lett. {\bf B702}, 429 (2011).

\bibitem{TST10}N.~Tajima, Y.~R.~Shimizu, and S.~Takahara, Phys. Rev. {\bf C82}, 034316 (2010).

\bibitem{Kruppa98}A.~T.~Kruppa, Phys. Lett.  {\bf B431}, 237 (1998).

\bibitem{OSTT10}
T.~Ono, Y.R.~Shimizu, N.~Tajima, S.~Takahara,
Phys. Rev. {\bf C82}, 034310 (2010).

\bibitem{TSS02}N.~Tajima, Y.~R.~Shimizu, and N.~Suzuki,
 Prog. Theor. Phys. Suppl. {\bf 146}, 628 (2002).

\bibitem{LW61} R.~H.~Lemmer and V.~F.~Weisskopf, Nucl. Phys. {\bf 25}, 624 (1961).

\bibitem{Ari04}
  K.~Arita,
  Int. J. Mod. Phys. E, {\bf 13}, 191 (2004).

\bibitem{HM09}I.~Hamamoto and B.~R.~Mottelson, Phys. Rev. {\bf C79},
 034317 (2009).

\bibitem{Zic85} 
  W. Zickendraht, Phys.\ Rev.\ Lett.\ {\bf 54}, 1906 (1985).

\bibitem{BDJ72}
 M.~Brack, J.~Damg{\aa}rd, A.~S.~Jensen, H.~C.~Pauli, V.~M.~Strutinsky,
  and C.~Y.~Wong, Rev. Mod. Phys. {\bf 44}, 320 (1972).

\bibitem{BR85} 
  T. Bengtsson and I. Ragnarsson, Nucl. Phys. {\bf A436},14 (1985).

\bibitem{BDN89}
 R.~Bengtsson, J.~Dudek, W.~Nazarewicz and P.~Olanders,
 Physica Scripta {\bf 39}, 196 (1989).

\bibitem{CDN87}S.~\'{C}wiok, J.~Dudek, W.~Nazarewicz, J.~Skalski, and
 T.~Werner, Comput. Phys. Commun. {\bf 46}, 379 (1987).

\bibitem{BW60}J.~Blomquist, S. Wahlborn, Ark, Fiz. {\bf 16}, 543 (1960).

\bibitem{Ro68}E.~Rost, Phys. Lett. {\bf B26}, 184 (1968).

\bibitem{Ch67}V. A.~Chepurnov, Yad. Fiz. {\bf 6}, 955 (1967).

\bibitem{BVC10}
 A.~Bhagwat, X.~Vi$\tilde{\rm n}$as, M.~Centelles, P.~Schuck, and R.~Wyss,
 Phys.\ Rev.\ C {\bf 81}, 044321 (2010).

\bibitem{WysPrivate05}
 R.~Wyss, private communication.

\bibitem{SS09} 
 T.~Shoji and Y.~R.~Shimizu, Prog.\ Theor.\ Phys.\ {\bf 121}, 319 (2009).

\bibitem{NWD94} 
W.~Nazarewicz, T.~R.~Werner, J.~Dobaczewski,
 Phys.\ Rev.\ C {\bf 50}, 2860, (1994).

\bibitem{Af99}A.~V.~Afanasjev, D.~B.~Fossan, G.~J.~Lane and I.~Ragnarsson, Phys. Rep. {\bf 322}, 1 (1999).

\bibitem{NDB85}
 W.~Nazarewicz, J.~Dudek, R.~Bengtsson, T.~Bengtsson, and I.~Ragnarsson,
 Nucl.\ Phys.\ {\bf A435}, 397 (1985).

\bibitem{TTO96} 
  N. Tajima, S. Takahara, and N. Onishi,
  Nucl.\ Phys.\ {\bf A603}, 23 (1996).

\end{thebibliography}
\end{document}